\documentclass[%
 reprint,floatfix,
 amsmath,amssymb,
 aps,
longbibliography
]{revtex4-2}

\usepackage{graphicx}% Include figure files
\usepackage{dcolumn}% Align table columns on decimal point
\usepackage{bm}% bold math
\usepackage[utf8]{inputenc}
\usepackage{textgreek}
\usepackage{subcaption}
\usepackage{booktabs}

\begin{document}

\preprint{APS/123-QED}

\title{Decoupling Structural Heterogeneity from Functional Fairness in Complex Networks:\\A Theoretical Framework based on the Imbalance Metric}

\author{Zhiyuan Ren}
\email{Corresponding author: zyren@xidian.edu.cn} 
\affiliation{Xidian University, Xi'an, China}

\author{Zhiliang Shuai}
\affiliation{Xidian University, Xi'an, China}

\author{Wenchi Cheng}
\affiliation{Xidian University, Xi'an, China}

\author{Kun Yang}
\affiliation{University of Essex, Colchester, UK}

\date{\today}

\begin{abstract}
Performance evaluation of complex networks has traditionally focused on structural integrity or average transmission efficiency, perspectives that often overlook the dimension of functional fairness. This raises a central question: Under certain conditions, structurally heterogeneous networks can exhibit high functional fairness. To systematically address this issue, we introduce a new metric, Network Imbalance ($I$), designed to quantitatively assess end-to-end accessibility fairness from a perceived  QoS perspective. By combining a tunable sigmoid function with a global Shannon entropy framework, the $I$ metric quantifies the uniformity of connection experiences between all node pairs. We analyze the mathematical properties of this metric and validate its explanatory power on various classical network models. Our findings reveal that low imbalance (i.e., high functional fairness) can be achieved through two distinct mechanisms: one via topological symmetry (e.g., in a complete graph) and the other via extreme connection efficiency driven by structural inequality (e.g., in a scale-free network). This decoupling of structure and function provides a new theoretical perspective for network performance evaluation and offers an effective quantitative tool for balancing efficiency and fairness in network design.
\end{abstract}

% \keywords{Network Imbalance}%Use showkeys class option if keyword
                              %display desired
\maketitle

\section{Introduction}

With the advancement of information technology, various complex network systems have permeated all levels of the social and economic fabric. The core function of these networks lies in the efficient and reliable transmission of information and resources. However, their inherent structural heterogeneity and dynamic evolutionary characteristics make the precise quantification and evaluation of overall network performance and stability a persistent theoretical and practical challenge.

Early research focused on structural inequality, often characterizing the heterogeneity of node connections through the degree distribution and its entropy or Gini coefficient \cite{ref1, ref2}, and identifying information flow bottlenecks using metrics like betweenness centrality \cite{ref3, ref4}. At a more formal level, methods based on automorphism groups \cite{ref5}, distance distribution functions \cite{ref6, ref7}, and graph kernel functions \cite{ref8} have been used to describe topological symmetry. Although these metrics can effectively identify structural features, they struggle to directly quantify the practical impact of such structural imbalances on the fairness of end-to-end communication experiences for all users across the network. To understand network complexity, information-theoretic methods have been widely introduced, using Shannon entropy to measure the randomness of network structure. The calculation objectives have evolved from the initial degree sequences \cite{ref9, ref10} and the betweenness distributions \cite{ref11} to approximations of the entropy of the von Neumann graph \cite{ref12} and distributions of higher-order motifs \cite{ref13}. However, the limitation of such methods is that they primarily aim to describe the structural randomness of the network, rather than directly quantifying the fairness of its service quality distribution when performing specific functions. Meanwhile, path-based analysis constitutes another major pillar of network science. From the classic average path length \cite{ref4} and the centrality of the betweenness \cite{ref11} to the more refined  morphospace of routing efficiency  \cite{ref14} and efficient path counting algorithms \cite{ref16, ref17}, the central role of paths in evaluating network efficiency has been emphasized. However, these methods typically compress path information into a single statistical value, losing distributional details, and lack a tunable mechanism to investigate connection performance sensitivity under different  QoS requirements. Finally, to link topology with specific network performance, cutting-edge work has utilized graph neural networks for QoS prediction \cite{ref18, ref19, ref20}, designed multi-constrained QoS routing algorithms \cite{ref21} and multiobjective topology optimization \cite{ref22}, and applied $\alpha$ fairness schemes to balance efficiency and fairness \cite{ref14, ref23}, or associated structure with dynamic performance through information diffusion analysis \cite{ref15} and deep reinforcement learning. Although these works confirm the critical impact of topology on performance, they are often a posteriori correlational analyses or decision models, failing to  implant  performance requirements (QoS) as an endogenous variable into the definition of the topological metric itself. Therefore, we face a challenge: How can we quantify the functional fairness of a network? This further leads to the core question this study will explore: is it possible for a structurally unequal system to achieve a high degree of functional fairness?

While traditional statistics like the variance of the shortest path distribution can measure the dispersion of path lengths, they are  context-blind  and cannot distinguish the practical meaning of  inequality  under different functional requirements. To address this challenge, this study proposes a new metric, Network Imbalance (I). In this paper, we define \textbf{ Functional Fairness } as the uniformity of the distribution of end-to-end connection experiences in a network. The imbalance metric is a quantitative evaluation tool designed for this purpose. It operates from a perceived QoS perspective through a two-step core mechanism: first, it transforms the cost of the path (in this case, hop count) between all node pairs into a connection experience evaluation score (weight w) in the [0,1] interval using a tunable Sigmoid function; second, it uses Shannon entropy to quantify the uniformity of the probability distribution formed by these evaluation scores. Ultimately, through the form $I=1-\text{normalized entropy}$, a high I value corresponds directly to a low entropy polarized distribution of connection experiences, i.e., functional unfairness. This parameterized design allows the Imbalance metric to serve as a flexible analytical tool, revealing the functional behavior of networks under different evaluation criteria.

Based on the ideas above, the main contributions of this paper can be summarized as follows:
\begin{enumerate}
    \item \textbf{Proposing and characterizing a new metric}: We propose the Network Imbalance (I), which introduces functional fairness as a core evaluation dimension alongside traditional efficiency and stability. It is founded on the principles that: (a) a network's health depends on the balance of its service distribution, avoiding the performance degradation associated with functional concentration; and (b) its evaluation must be adaptive to business needs through a context-aware QoS lens."
    \item \textbf{Analyzing its mathematical foundation}: We systematically analyze the mathematical properties of the imbalance metric, including its boundedness, continuity, differentiability, and extreme value conditions.
    \item \textbf{Conducting a coupling analysis with classical metrics}: We explore the relationship between the Imbalance metric and existing network metrics, revealing the potential decoupling mechanism between structural heterogeneity and functional homogeneity.
    \item \textbf{Studying its behavior in random graph models}: We systematically study the behavioral characteristics of the Imbalance metric in ER, BA, and WS models, demonstrating its ability to capture network phase transitions and structural properties.
    \item \textbf{Exploring its application in topology optimization}: We propose a network topology optimization method based on the Imbalance metric and verify its feasibility in guiding network structure adjustments to enhance global functional efficiency.
\end{enumerate}

The remainder of this paper is organized as follows. Section II elaborates on the theory and mathematical properties of the Network Imbalance metric. Section III presents the simulation and theoretical analysis results for various network models. Section IV provides a detailed discussion of the physical significance and application value of the research findings. Finally, Section V concludes the entire paper.

\section{The Network Imbalance Metric: Theory and Properties}

\subsection{Fundamental Graph-Theoretic Notation}
Before elaborating on the calculation method of the Imbalance metric, this section first introduces some fundamental graph-theoretic notations and definitions that will be used in the subsequent discussion. Consider an undirected graph $G=(V,E)$, where $V$ is the set of nodes and $E$ is the set of edges. The total number of nodes in the network is denoted as $N = |V|$, and the total number of edges is $M = |E|$, with $N \ge 2$. For any two distinct nodes $u, v \in V$, their shortest hop count, denoted as $h(u,v)$, is defined as follows:
\begin{equation}
       h(u,v) =
      \begin{cases}
        \min\{ \text{length}(P_{uv}) \}, & \text{if } u \text{ is reachable from } v \\
        +\infty, & \text{otherwise}
      \end{cases}
\end{equation}
where $P_{uv}$ represents a path from node $u$ to node $v$, and $\text{length}(P_{uv})$ is the number of edges (hops) on this path. When nodes $u$ and $v$ are unreachable, their shortest hop count is considered infinite.

\subsection{Formal Definition of the Imbalance Metric}
The core idea of the Imbalance metric is to map the shortest path lengths between all node pairs to connection weights via a nonlinear sigmoid function, then calculate the normalized Shannon entropy of the distribution of these connection weights, and finally obtain a scalar value reflecting the functional fairness of the network.

In the theoretical construction part of this study, we first choose the most fundamental, unweighted shortest hop count as the distance metric. This choice is intended to isolate the variables, allowing us to first focus on the functional imbalance caused purely by the topological structure itself, without interference from heterogeneous link costs. This selection provides a clear, controllable theoretical baseline for the analysis of the basic mathematical properties of the imbalance metric and its behavior in classical random graph models. The extension of this framework to more general weighted networks will be discussed in detail in Section V-C.

The effective hop count $d(u,v)$ between any pair of nodes $(u,v)$ is its shortest hop count $h(u,v)$:
\begin{equation}
    d(u,v) = h(u,v)
\end{equation}
Since the shortest hop count $h(u,v)$ is a discrete integer, to construct a continuous function that is easy to analyze and to introduce a service-oriented sensitivity, we smooth $h(u,v)$ using a Sigmoid function. The Sigmoid function not only converts discrete hop values into continuous weights but also has the desirable mathematical property of being infinitely differentiable. At the same time, it naturally handles cases where the hop count is infinite and allows for the introduction of a tunable parameter $h_0$ as a benchmark for quality of service.

We introduce two tunable parameters: $h_0 > 0$ (ideal hop count threshold) and $a > 0$ (sigmoid steepness). For each pair of nodes $(u,v)$, its sigmoid-adjusted weight $w(u,v)$ is defined as:
\begin{equation}
    w(u,v) =
    \begin{cases}
        \frac{1}{1 + \exp[a(d(u,v) - h_0)]}, & \text{if } d(u,v) < +\infty \\
        0, & \text{if } d(u,v) = +\infty
    \end{cases}
\end{equation}
When $d(u,v) \ll h_0$, $w(u,v) \approx 1$; when $d(u,v) \gg h_0$, $w(u,v) \approx 0$.

\begin{figure}[h!]
 \includegraphics[width=\columnwidth]{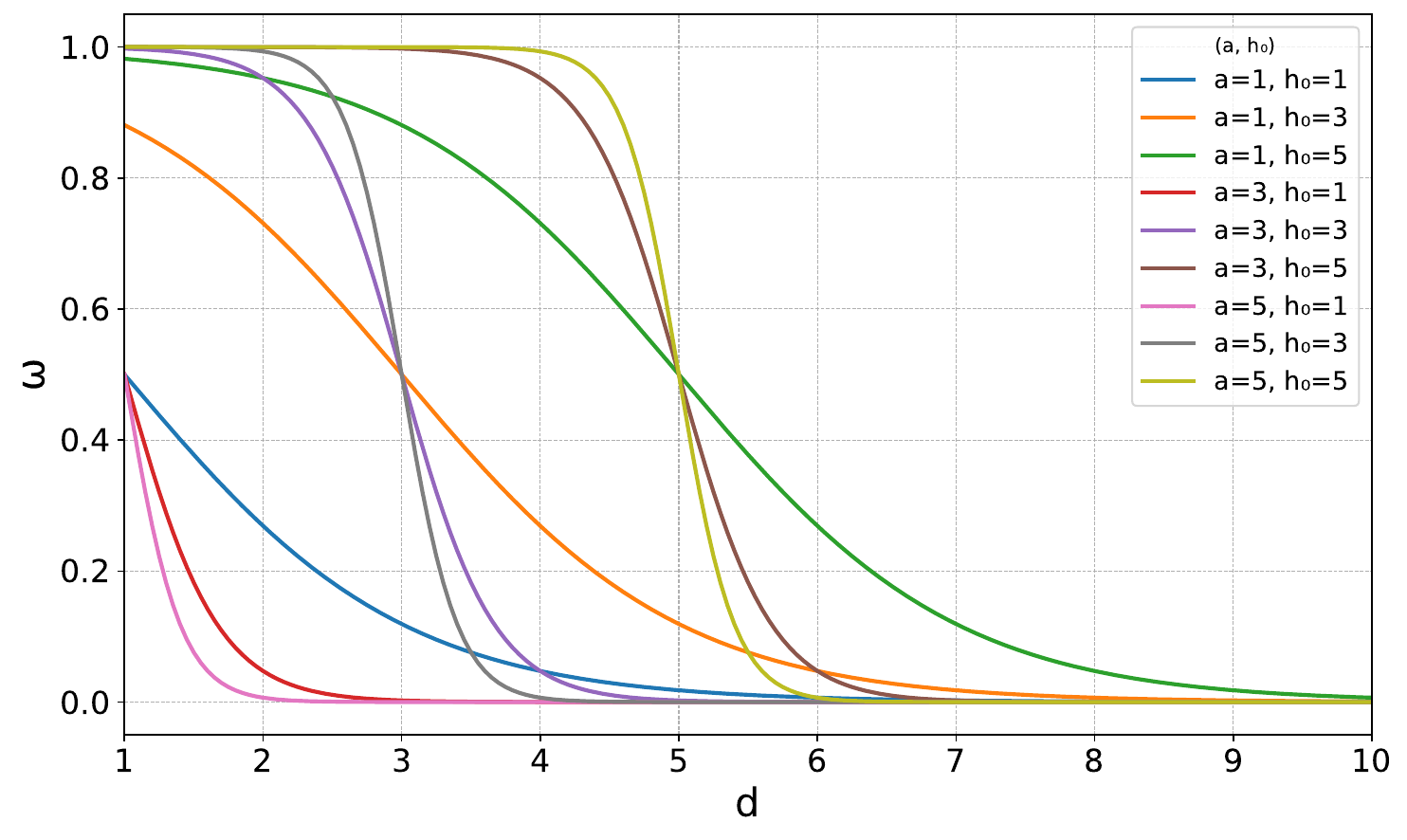}
 \caption{\label{fig:params} The relationship between connection weight $w(u,v)$ and parameters $a, h_0$.}
\end{figure}

 First, we calculate the total weight of the network, $W = \sum_{x \neq y} w(x,y)$.

 We must consider the edge case where the graph has no reachable paths (e.g., an empty graph $E=\emptyset$), resulting in $W=0$. In this scenario, the normalization is ill-defined. We explicitly define the Imbalance $I=1$ in such cases, representing the state of maximum functional imbalance.
 If $W>0$, we normalize the weights $w(u,v)$ for all ordered node pairs $(u,v)$ (where $u \neq v$) to construct a probability distribution $\{p_{u,v}\}$:
 \begin{equation}
    p_{u,v} = \frac{w(u,v)}{W}
 \end{equation}
 
 such that $\sum_{u \neq v} p_{u,v} = 1$. This yields the normalized probability distribution $p_{u,v}$. At this point, $p(u,v)$ is a discrete probability distribution, for which we can define the normalized Shannon entropy $Q$:
\begin{equation}
    H = - \sum_{u \neq v} p_{u,v} \log_2(p_{u,v})
\end{equation}
\begin{equation}
    Q = \frac{H}{\log_2(N(N-1))}
\end{equation}
where $N(N-1)$ is the total number of ordered node pairs in the network. $Q \in [0,1]$. When $w(u,v)$ of all pairs of accessible nodes are equal, $Q=1$.

Finally, we provide the definition of imbalance for the unconstrained case of a homogeneous network. The imbalance metric is ultimately defined as the complement of the normalized Shannon entropy $Q$:
\begin{equation}
     I = 1 - Q
\end{equation}
Therefore, $I \in [0,1]$. $I \approx 0$ indicates a uniform distribution of network weights and good connectivity; $I \approx 1$ indicates a highly concentrated distribution of network weights and poor or unbalanced connectivity.

\subsection{Computational Complexity Analysis}
Calculating network imbalance $I(G; a, h_0)$ involves mainly two core steps: first, determining the shortest path lengths between all pairs of nodes in the network; second, performing a weighted summation and normalization based on these path lengths and their weights.

The first step, calculating the lengths of the shortest paths $d(u,v)$ between all pairs of nodes $(u,v)$ (that is, the problem of the shortest path of all pairs, APSP), is the main bottleneck that determines the total computational complexity. For a graph $G=(V, E)$ with $N$ nodes and $M$ edges:
\begin{itemize}
    \item For a \textbf{dense graph} ($M \approx O(N^2)$), the Floyd-Warshall algorithm is typically used, with a time complexity of $O(N^3)$.
    \item For a \textbf{sparse graph} ($M \approx O(N)$ or $O(N \log N)$), a more efficient method is to run the Dijkstra algorithm for each node. The time complexity of a single Dijkstra run is $O(M + N \log N)$, so the total time complexity for all source nodes $N$ is $O(N(M + N \log N))$. In extremely sparse graphs (such as path or star graphs), $M=O(N)$, and the complexity can be simplified to $O(N^2 \log N)$.
\end{itemize}

The second step, once all the shortest path lengths $d(u,v)$ are known, involves calculating $w(u,v)$ and performing a summation for all ordered node pairs $N(N-1)$. The time complexity of this process is $O(N^2)$.

In summary, the overall computational complexity of the imbalance metric is determined primarily by the APSP step. For dense graphs, the complexity is $O(N^3)$; for sparse graphs, it is $O(N(M + N \log N))$. This implies that for very large-scale networks, the exact calculation of the imbalance metric faces significant computational challenges.

\subsection{Relationship with the Hop-count Distribution}
The imbalance metric has the most direct and fundamental connection with the network's Hop-count Distribution. In fact, imbalance can be seen as a complex statistical functional acting on the set of shortest path hop counts of all pairs $\{h(u,v)\}$, designed to quantify the  non-uniformity  of the entire hop count distribution after a non-linear mapping.

Let $N_h$ be the number of pairs of nodes in the network whose shortest path is $h$. The probability mass function of the hop count distribution can be defined as $P(h) = \frac{N_h}{N(N-1)}$. The calculation of imbalance is based on this distribution:
\begin{enumerate}
    \item  \textbf{From Hop-count Distribution to Weight Distribution}: Each hop count $h$ is assigned a weight $w(h) = (1 + e^{a(h-h_0)})^{-1}$ using the Sigmoid function. Thus, the total weight of the network $W$ can be obtained by a weighted sum over $P(h)$:
        \begin{equation}
        W = \sum_{h=1}^{\text{diam}(G)} N_h \cdot w(h) = N(N-1) \sum_{h=1}^{\text{diam}(G)} P(h) \cdot w(h)
        \end{equation}
    where $\text{diam}(G)$ is the diameter of the graph.

    \item  \textbf{From Weight Distribution to Entropy}: All pairs of nodes with the same hop count $h$ will have the same normalized probability $p_{u,v}$, denoted $p(h) = w(h)/W$. Therefore, the total Shannon entropy $H$ of the network can also be expressed in terms of $P(h)$:
        \begin{equation}
        H = - \sum_{h=1}^{\text{diam}(G)} N_h \cdot p(h) \log_2(p(h))
        \end{equation}
        which can be rewritten as \begin{equation}
        H = - N(N-1) \sum_{h=1}^{\text{diam}(G)} P(h) \cdot \frac{w(h)}{W} \log_2\left(\frac{w(h)}{W}\right)
        \end{equation}
\end{enumerate}
From the mathematical relationships above, it is clear that the  Imbalance  metric does not simply respond to an isolated feature of the hop count distribution $P(h)$, but is a complex mapping of its entire form. Although traditional metrics such as average path length (the first moment of $P(h)$) and network diameter (the upper bound of the support of $P(h)$) provide a macroscopic assessment of network efficiency, they cannot fully determine the value of  Imbalance . The  I  metric depends more deeply on the second- and higher-order statistical properties of the distribution, such as its variance (characterizing the width of the distribution) and skewness (characterizing the symmetry of the distribution).

An extremely concentrated $P(h)$ will inevitably lead to $I=0$, while a $P(h)$ spread over a wide range tends to produce a higher $I$ value. More importantly, the value of Imbalance also depends on the interaction between the fine-grained shape of $P(h)$ and the evaluation parameters $h_0$ and $a$. This pair of parameters acts as an adjustable observation lens, through which we can probe the functional fairness of the network at different connection scales. Therefore, Imbalance provides a comprehensive perspective, measuring \textbf{ fairness and balance} of the connection experience—as determined by the entire hop count distribution $P(h)$ and filtered through a specific QoS lens across all node pairs in the network. This explains why two networks with identical average path lengths may have vastly different  imbalance   values due to differences in the variance or skewness of their path distributions, or due to their performance differences under specific service requirements.

\subsection{Fundamental Mathematical Properties of Imbalance}
This section is dedicated to a comprehensive mathematical analysis of the Network Imbalance metric to reveal its intrinsic structural and functional characteristics. We will successively discuss the metric's boundedness (ensuring its value range is meaningful), continuity (ensuring a smooth response to small network changes), differentiability (providing a theoretical basis for gradient-based optimization methods), and derive its extreme value conditions under specific circumstances. The elucidation of these fundamental mathematical properties will provide a solid theoretical foundation for the robustness and universality of the imbalance metric as a new network evaluation tool.

\subsubsection{Existence and Value Range}

\textbf{Proposition 2.1} For any graph $G=(V, E)$ with $N$ nodes ($N \ge 2$), the range of values of its imbalance metric is always $[0, 1]$.
\begin{equation}
0 \le I(G) \le 1
\end{equation}

\textbf{Proof:} This proposition follows directly from the definition of the imbalance metric.
\begin{enumerate}
    \item \textbf{Properties of normalized weights $p_{u,v}$}: By definition, $p_{u,v} = \frac{w(u,v)}{\sum_{x \neq y} w(x,y)}$. Since the weight function $w(u,v) = \frac{1}{1 + e^{a(d(u,v) - h_0)}}$ is always positive ($w(u,v) > 0$), and the denominator is a finite positive value (for a graph with $N \ge 2$, there is at least one pair of nodes), it follows that $p_{u,v} > 0$. By definition, the sum of all $p_{u,v}$ is 1: $\sum_{u \neq v} p_{u,v} = 1$. Thus, $\{p_{u,v}\}$ constitutes a valid discrete probability distribution.
    \item \textbf{Bounds of Shannon entropy $H$}: The Shannon entropy is defined as $H = - \sum_{u \neq v} p_{u,v} \log_2(p_{u,v})$. According to the fundamental principles of information theory, for a probability distribution with $K$ possible events, its Shannon entropy $H$ is bounded by $0 \le H \le \log_2(K)$. In this study, $K$ is the total number of pairs of ordered nodes, that is, $K = N(N-1)$. Therefore, the range of Shannon entropy $H$ is $0 \le H \le \log_2(N(N-1))$. $H_{min} = 0$ if and only if $p_{u,v}$ is a deterministic distribution (that is, one $p_{u,v}=1$, and the rest are 0). $H_{max} = \log_2(N(N-1))$ if and only if $p_{u,v}$ is a uniform distribution (that is, all $p_{u,v} = \frac{1}{N(N-1)}$).
    \item \textbf{Bounds of normalized Shannon entropy $Q$}: The normalized Shannon entropy is defined as $Q = \frac{H}{\log_2(N(N-1))}$. Since $H \ge 0$ and $\log_2(N(N-1)) > 0$ (for $N \ge 2$), it follows that $Q \ge 0$. Combined with $H \le \log_2(N(N-1))$, we get $Q = \frac{H}{\log_2(N(N-1))} \le \frac{\log_2(N(N-1))}{\log_2(N(N-1))} = 1$. Thus, the range of values of the normalized Shannon entropy $Q$ is $0 \le Q \le 1$.
    \item \textbf{Bounds of the Imbalance metric}: The final definition of the imbalance metric is $I = 1 - Q$. When $Q$ takes its maximum value $Q_{max} = 1$, $I_{min} = 1 - 1 = 0$. This corresponds to a uniform distribution of $p_{u,v}$, that is, the most uniform state of the weights of the network connection. When $Q$ takes its minimum value $Q_{min} = 0$, $I_{max} = 1 - 0 = 1$. This corresponds to a deterministic distribution of $p_{u,v}$, i.e. an extremely unbalanced state of the network connection weights.
\end{enumerate}
In conclusion, for any graph $G=(V, E)$, its Imbalance metric necessarily exists and its value lies within the $[0, 1]$ interval. Q.E.D.

\subsubsection{Continuity and Differentiability}
This section analyzes the response characteristics of the Imbalance metric to changes in its inputs, including the network topology and model parameters. We will prove that the Imbalance metric is a continuous function and, under certain conditions, differentiable, which makes it a robust metric insensitive to minor perturbations and provides possibilities for gradient-based network optimization algorithms.

\paragraph{Continuity with respect to Model Parameters}
In this section, our aim is to mathematically prove that for a given network topology, the imbalance metric is a well-behaved function of its model parameters $a$ and $h_0$. We will state its continuity through a rigorous proposition, which is crucial for ensuring the stability of the metric and for subsequent gradient-based analysis.

\textbf{Proposition 2.2} For any given graph $G=(V, E)$ with $N \ge 2$ nodes, the imbalance metric is a continuous function of its model parameters $a \in \mathbb{R}$ and $h_0 \in \mathbb{R}$.

\textbf{Proof:} Our proof strategy is to decompose the calculation of the imbalance metric into a chain of function compositions and demonstrate the continuity of each link in the chain. The entire calculation process can be abstracted as:
$(a, h_0) \xrightarrow{f_1} \{w_{u,v}\} \xrightarrow{f_2} \{p_{u,v}\} \xrightarrow{f_3} H \xrightarrow{f_4} I$

\begin{enumerate}
    \item \textbf{Continuity of the weight function $w_{u,v}$}: For any fixed pair of nodes $(u,v)$ in the graph, their shortest path $h(u,v)$ is a constant. Therefore, the connection weight function $w(u,v; a, h_0) = (1 + \exp[a(h(u,v) - h_0)])^{-1}$ is made up of linear, exponential and reciprocal functions. These are all elementary functions and are continuous everywhere in their respective domains. Thus, as a function of variables $a$ and $h_0$, $w(u,v; a, h_0)$ is continuous throughout the $\mathbb{R}^2$ space.
    \item \textbf{Continuity of the total $W$ and probability distribution $\{p_{u,v}\}$}: The total weight of all ordered node pairs in the network, $W(a, h_0) = \sum_{u \neq v} w(u,v; a, h_0)$, is the sum of a finite number of continuous functions and therefore also a continuous function of $(a, h_0)$. For a non-empty graph ($N \ge 2$) with at least one edge, there is at least one finite hop count $h(u,v)$, ensuring that there is at least one weight $w(u,v)>0$. Thus, the total weight $W(a, h_0)$ is always positive. The normalized probability $p_{u,v}(a, h_0) = \frac{w(u,v; a, h_0)}{W(a, h_0)}$ is the quotient of two continuous functions with a non-zero denominator, so $p_{u,v}$ is also a continuous function of $(a, h_0)$.
    \item \textbf{Continuity of Shannon entropy $H$}: Shannon entropy $H = - \sum_{u \neq v} p_{u,v} \log_2(p_{u,v})$ is a continuous function of the probability distribution vector $\mathbf{p} = \{p_{u,v}\}$ (in information theory, it is conventional to set $0 \log 0 = 0$). Since the mapping from the parameter space $(a, h_0)$ to the probability distribution space $\{\mathbf{p}\}$ is continuous (as proved in the previous step), and the mapping from the probability distribution space to the entropy value $H$ is also continuous, by the composite function continuity theorem, the Shannon entropy $H$ must be a continuous function of the original parameters $(a, h_0)$.
    \item \textbf{Continuity of the Imbalance metric}: Finally, the imbalance metric is defined by $I = 1 - \frac{H}{\log_2(N(N-1))}$. This is a linear transformation of the continuous function $H$. Since linear transformations preserve continuity, we conclude that the imbalance metric is a continuous function of its model parameters $a$ and $h_0$.
\end{enumerate}
Q.E.D.

It must be emphasized that the continuity in the above proposition is with respect to the model parameters. A change in network topology, such as adding or removing an edge, is a discrete event. This event causes a discontinuous jump in the underlying shortest path vector $\{h(u,v)\}$. Therefore, the imbalance metric itself is not continuous with respect to changes in network topology. However, the metric exhibits good robustness to such discrete changes. Thanks to the smoothing property of the Sigmoid function, a small integer jump in one or more values of $h(u,v)$ (for example, from 3 to 2) will only cause a finite, bounded, smooth change in the corresponding weight $w(u,v)$. Unless the addition or removal of the edge leads to a fundamental change in the macroscopic connectivity of the network (e.g., splitting a connected graph into two disconnected components), a local topological perturbation will not cause a catastrophic change in the global imbalance value. This robustness ensures that the results of Imbalance as an evaluation tool are stable and meaningful for the minor evolution of networks.

\paragraph{Differentiability with respect to Model Parameters}
Following the proof of continuity, we further investigate the differentiability of the imbalance metric.

\textbf{Proposition 2.3} For any given graph $G=(V, E)$ with $N \ge 2$ nodes, the imbalance metric is a differentiable function of its model parameters $a \in \mathbb{R}$ and $h_0 \in \mathbb{R}$.

\textbf{Proof:} We will demonstrate that each component in the function chain that constitutes the imbalance metric is differentiable.
\begin{enumerate}
    \item \textbf{Differentiability of the weight function $w_{u,v}$}: The weight function $w(u,v; a, h_0)$ is a standard Sigmoid function, composed of elementary functions. It is well-known that the Sigmoid function is infinitely differentiable (a $C^\infty$ function) over its entire domain. Therefore, for a fixed $h(u,v)$, $w(u,v; a, h_0)$ is differentiable with respect to its variables $a$ and $h_0$. Its partial derivatives can be analytically computed.
    \item \textbf{Differentiability of the total weight $W$ and the probability distribution $\{p_{u,v}\}$}: The total weight $W(a, h_0)$ is the sum of a finite number of differentiable functions $w(u,v; a, h_0)$, and by the linear differentiation rule, $W(a, h_0)$ is also a differentiable function of $(a, h_0)$. Consequently, the probability $p_{u,v} = w_{u,v} / W$ is the quotient of two differentiable functions, with the denominator $W$ being always non-zero. According to the quotient rule, $p_{u,v}$ is also a differentiable function of $(a, h_0)$.
    \item \textbf{Differentiability of Shannon entropy $H$}: The Shannon entropy function $H(\mathbf{p}) = -\sum p_i \log_2 p_i$ is differentiable in its domain (i.e. inside the probability simplex, where all $p_i > 0$). In our model, for any finite parameters $(a, h_0)$, all weights $w(u,v)$ are strictly positive, and hence all probabilities $p_{u,v}$ are also strictly positive, ensuring that the function $H(\mathbf{p})$ is in the differentiable region. By the multivariate chain rule, since $H$ is differentiable with respect to its parameters $\{p_{u,v}\}$, and each $p_{u,v}$ is in turn differentiable with respect to $(a, h_0)$, the composite function $H(a, h_0)$ must be differentiable with respect to $(a, h_0)$.
    \item \textbf{Differentiability of the Imbalance metric}: Since the imbalance metric is a simple linear transformation of the differentiable function $H$, it naturally inherits differentiability.
\end{enumerate}
Q.E.D.

The differentiability established by Proposition 2.3 shows that the imbalance metric can be analyzed as an objective function or a constraint term. We can analytically calculate the partial derivatives of the imbalance metric with respect to the parameters $a$ and $h_0$, thus obtaining the gradient vector $\nabla I(a, h_0) = \left( \frac{\partial I}{\partial a}, \frac{\partial I}{\partial h_0} \right)$. This gradient precisely indicates the direction in the parameter space in which the imbalance value increases most rapidly. Therefore, for a given network topology, we can use optimization algorithms to automatically search for a set of parameters $(a^*, h_0^*)$ to maximize or minimize the imbalance value of the network. This process can not only profoundly reveal under which  service perspective  a specific network appears most ordered or most chaotic, but can also actively guide analysis and design as an operable and programmable analytical metric.

\subsubsection{Analysis of Extreme Values}

This section will rigorously prove the conditions under which the Imbalance metric reaches or approaches its minimum (0) and maximum (1) values, respectively, under various topological and parameter conditions, thereby providing a theoretical cornerstone for understanding  completely ordered  and  extremely disordered  network states.

\paragraph{The State of Minimum Imbalance}
We first investigate the conditions for an imbalance value of 0, which represents a functionally most balanced network state.

\textbf{Proposition 2.4} For a connected graph $G$ and any finite non-zero parameter $a$, a necessary and sufficient condition for $I(G) = 0$ is that the shortest path lengths $h(u,v)$ between all pairs of nodes in the graph are equal.

\textbf{Proof:}
($\Rightarrow$) \textbf{Necessity}: Assume $I(G) = 0$.
\begin{enumerate}
    \item By definition, $I(G) = 1 - Q(G) = 0$ implies that the normalized entropy $Q(G) = 1$.
    \item $Q(G) = H / H_{max}$, so the entropy $H$ must reach its theoretical maximum value $H_{max} = \log_2(N(N-1))$.
    \item According to the fundamental principles of information theory, a necessary and sufficient condition for the Shannon entropy to be maximized is that the corresponding probability distribution is uniform. That is, for all $u \neq v$, $p_{u,v}$ must be a constant: $p_{u,v} = \frac{1}{N(N-1)}$.
    \item Since $p_{u,v} = w_{u,v} / W$, for all $p_{u,v}$ to be equal, it is necessary and sufficient that all connection weights $w_{u,v}$ are equal.
    \item The weight function $w_{u,v} = (1 + \exp[a(h(u,v) - h_0)])^{-1}$ is a strictly monotonic function of its argument $h(u,v)$ (because $a \neq 0$). Therefore, for all $w_{u,v}$ to be equal, it is necessary and sufficient that all shortest path lengths $h(u,v)$ are equal.
\end{enumerate}

($\Leftarrow$) \textbf{Sufficiency}: Assume that all $h(u,v)$ are equal to a constant $h_c$.
\begin{enumerate}
    \item Then, all weights $w_{u,v} = (1 + \exp[a(h_c - h_0)])^{-1}$ must be equal, let us denote it as $w_c$.
    \item The total weight $W = \sum_{u \neq v} w_c = N(N-1)w_c$.
    \item Therefore, each probability $p_{u,v} = w_c / W = \frac{1}{N(N-1)}$, forming a uniform distribution.
    \item The entropy $H$ of this uniform distribution must be the maximum entropy $H_{max}$.
    \item Thus, $Q(G)=1$, which finally yields $I(G)=0$.
\end{enumerate}

\textbf{Corollary 2.1} In a non-trivial ($N>2$) connected graph, the only topological structure that can make the imbalance value 0 is the \textbf{Complete Graph ($K_N$)}.

\textbf{Proof:} Proposition 2.4 requires the shortest path length between all pairs of nodes to be equal. In a connected graph, this is only possible if all nodes are directly adjacent, that is, all $h(u,v)=1$. This is precisely the definition of a complete graph $K_N$.

\paragraph{The State of Maximum Imbalance}
Unlike the minimum value, the maximum value of 1 for Imbalance is a supremum that cannot be physically reached but can be approached indefinitely. This situation represents the extreme imbalance of the functional distribution of the network.

\textbf{Proposition 2.5} For any graph $G$ and finite parameters, $I(G) < 1$. However, 1 is its least upper bound, that is, $\sup_{G, a, h_0} I(G) = 1$.

\textbf{Proof:}
\begin{enumerate}
    \item \textbf{Proof that $I(G) < 1$}: $I(G) = 1$ is equivalent to $H=0$. The only condition for zero entropy is that the probability distribution is a Kronecker delta function, ie, there exists a specific pair of nodes $(i,j)$ such that $p_{i,j}=1$, and all other $p_{u,v}=0$. This would require $w_{i,j} > 0$, while all the other $w_{u,v}=0$. However, the range of the weight function $w_{u,v} = (1 + \exp[\cdot])^{-1}$ is $(0, 1)$. For any finite $a, h_0,$ and $h(u,v)$, the weight value is always positive. Therefore, no weight can be zero, which means $H>0$, and thus $I(G)$ is always less than 1.
    \item \textbf{Proof that $\sup I(G) = 1$}: To prove that 1 is its supremum, we must construct a sequence of graphs or parameters for which the Imbalance value can arbitrarily approach 1. This is equivalent to making the entropy $H$ arbitrarily approach 0. Our strategy is to choose a specific topology and parameters to make the connection weights of a very small number of node pairs much greater than all other node pairs, thus making the probability distribution highly concentrated. Consider a graph $G$ where the set of shortest path lengths has a separable minimum. For example, there are pairs of $k$ nodes with a shortest path of $h_{min}$, and all other pairs of nodes have a shortest path of at least $h_{min} + \delta$ (where $\delta > 0$). For an unweighted graph, $\delta \ge 1$. A simple example is the \textbf{Path Graph ($P_N$)}, where only two pairs of neighboring nodes (a total of $k=2$ ordered pairs) have the shortest path $h_{min}=1$.
    Now, we set parameters to  amplify  this difference. Let $h_0 = h_{min} + \delta/2$, and consider the limit as the sharpness parameter $a \to \infty$:
    \begin{itemize}
        \item For the $k$ node pairs with the shortest path $h_{min}$, their weight $w \to (1 + \exp[a(h_{min} - (h_{min}+\delta/2))])^{-1} = (1+e^{-\infty})^{-1} \to 1$.
        \item For all other pairs of nodes with the shortest path $h  \ge h_{min}+\delta$, their weight $w  \to (1 + \exp[a(h  - (h_{min}+\delta/2))])^{-1} \to (1+e^{+\infty})^{-1} \to 0$.
    \end{itemize}
    In this limit, the total weight $W \to k \times 1 = k$. The probability distribution is almost entirely concentrated on these pairs of $k$ nodes, forming a uniform distribution over the $k$ states. Its entropy $H \to \log_2(k)$. Thus, the Imbalance value approaches:
    \begin{equation} \label{eq:your_label_here}
         \lim_{a \to \infty} I(G) = 1 - \frac{\log_2(k)}{\log_2(N(N-1))}
    \end{equation}
    To maximize this value (i.e., get closest to 1), we need to minimize the numerator $\log_2(k)$. In a connected graph, the minimum value of $k$ is 2 (that is, a pair of directly connected nodes $(u,v)$ and $(v,u)$). Substituting $k=2$, we get:
    \begin{equation} \label{eq:sup_information}
    \sup I(G) = 1 - \frac{1}{\log_2(N(N-1))}
    \end{equation}
    Clearly, as $N \to \infty$, this limit approaches 1. This shows that we can construct cases where the imbalance value is arbitrarily close to 1.
\end{enumerate}

\textbf{Example: Star graph ($S_N$)} The star graph is a classic example that approaches the maximum imbalance. It has $2(N-1)$ paths of length 1 (between the center and the periphery) and $(N-1)(N-2)$ paths of length 2 (between peripheral nodes). By setting $h_0 = 1.5$ and letting $a \to \infty$, we can concentrate all the weight on the length 1 paths, making $k=2(N-1)$. Its limiting imbalance is $1 - \frac{\log_2(2(N-1))}{\log_2(N(N-1))}$, which also approaches 1 as $N$ becomes large.

\subsubsection{Discussion on Monotonicity}

\paragraph{Monotonicity Analysis with respect to Model Parameters $a$ and $h_0$}
After determining the extreme limits of the metric, we further investigate its sensitivity to the model parameters $a$ (sharpness) and $h_0$ (threshold). This analysis is crucial because it reveals how we can  probe  a network s structural properties by adjusting this pair of parameters. We will show that the relationship between Imbalance and these two parameters is not simply monotonic but exhibits a more complex response pattern that depends on the global topology of the network. Our analysis will be based on studying the partial derivatives of the imbalance with respect to $a$ and $h_0$. From the chain rule, we have:
\begin{equation} \label{eq:partial_derivative_I}
\frac{\partial I}{\partial \theta} = - \frac{1}{H_{max}} \frac{\partial H}{\partial \theta} \quad (\text{where } \theta \in \{a, h_0\})
\end{equation}

Thus, the direction of change of Imbalance is opposite to the direction of change of the Shannon entropy $H$. Our core task is to analyze how changes in the parameters affect entropy $H$.

\textbf{Effect of parameter $a$ (sharpness)} Parameter $a$ controls the  steepness  of the Sigmoid evaluation function and can be understood as a  judgment stringency  or  contrast  knob.
\begin{itemize}
    \item \textbf{Mathematical analysis}: The partial derivative of the entropy $H$ with respect to $a$ is a complex process, but its sign ultimately depends on the weighted effect of the terms $\frac{\partial w_{u,v}}{\partial a}$. The partial derivative of the weight function with respect to $a$ is:
    \begin{equation} \label{eq:partial_w_a}
    \frac{\partial w_{u,v}}{\partial a} = - (h(u,v) - h_0) \cdot w_{u,v} (1-w_{u,v})
    \end{equation}
    where $w_{u,v}(1-w_{u,v}) > 0$. Therefore, the sign of the derivative is completely determined by $-(h(u,v)-h_0)$.
    \begin{itemize}
        \item When $h(u,v) < h_0$ (the path is  better  than the threshold), $\frac{\partial w_{u,v}}{\partial a} > 0$.
        \item When $h(u,v) > h_0$ (the path is  worse  than the threshold), $\frac{\partial w_{u,v}}{\partial a} < 0$.
    \end{itemize}
    \item \textbf{Physical interpretation and conclusion}: The above analysis reveals the central role of the parameter $a$: When $a$ increases (assuming $a>0$), it amplifies the difference between the length of the path $h(u,v)$ and the threshold $h_0$. It makes the weight of  good  paths ($h < h_0$) closer to 1, and the weight of  bad  paths ($h > h_0$) closer to 0. How this  stretching  effect influences the overall entropy $H$ depends on the original shape of the network hop-count distribution.
    \begin{enumerate}
        \item If the weights of most paths are already differentiated (e.g., most weights are high, a few are low), increasing $a$ will exacerbate this imbalance, making the weight distribution more  peaked,  leading to a decrease in entropy $H$. In this case, $\frac{\partial I}{\partial a} > 0$.
        \item In some special cases, if the network path weights are distributed over two or more  humps,  increasing $a$ might flatten these  humps,  making the overall distribution more uniform, leading to an increase in entropy $H$. In this case, $\frac{\partial I}{\partial a} < 0$.
    \end{enumerate}
    Although the latter possibility exists, in most application scenarios, increasing parameter $a$ tends to enhance the inherent inequality of connection quality in the network, thus leading to an increase in the Imbalance value. The parameter $a$ allows us to see the functional heterogeneity of the network more clearly under a given standard $h_0$.
\end{itemize}

\textbf{Effect of parameter $h_0$ (threshold)} Parameter $h_0$ defines the  passing grade  for acceptable path lengths and can be seen as a  tolerance  scale.
\begin{itemize}
    \item \textbf{Mathematical analysis}: The partial derivative of the weight function with respect to $h_0$ is:
    \begin{equation} \label{eq:partial_w_h0}
    \frac{\partial w_{u,v}}{\partial h_0} = a \cdot w_{u,v} (1-w_{u,v})
    \end{equation}
    \item \textbf{Physical interpretation and conclusion}: Assuming that we usually take $a>0$, then $\frac{\partial w_{u,v}}{\partial h_0}$ is always positive. This means that increasing the tolerance threshold $h_0$ will unidirectionally increase the connection weights $w_{u,v}$ of all paths. Intuitively, this is because a higher  passing grade  makes more paths be evaluated as  good paths.  However, the simultaneous increase of all weights has a complex effect on the entropy of the general distribution $H$, which depends on the original state of the distribution before the weight increase. Pushing all values towards 1 might make a previously uniform distribution non-uniform (entropy decreases), or it might make a previously very non-uniform distribution slightly more uniform (entropy increases). Therefore, the imbalance also does not have global monotonicity with respect to $h_0$. Its trend depends on the network hop-count distribution ${h(u,v)}$. Adjusting $h_0$, we can probe at which connection scale a network is most sensitive, and at which  passing grade  its connection balance is most easily  disrupted. 
\end{itemize}
Therefore, the complex, nonmonotonic relationship between Imbalance and its parameters indicates that Imbalance is a sensitive analytical tool that can be used to finely probe the multiscale structural features of a network through parameter adjustment.

\paragraph{Non-Monotonicity with Respect to Edge Addition}
A key question in network science is whether adding edges to a network (i.e., enhancing connectivity) always improves its functional fairness. Traditional efficiency metrics, such as the average path length, improve monotonically with the addition of edges. However, the response mechanism of the Imbalance metric, as a measure of the uniformity of the global connection experience distribution, is more complex.

\textbf{Proposition 2.6 (Non-monotonicity with respect to edge addition/deletion)} The imbalance metric is not a monotonic function of the set of edges. That is, for a graph $G=(V, E)$ and a new edge $e \notin E$, in the new graph $G =(V, E \cup \{e\})$, its imbalance value may change compared to that of the original graph $G$.

\textbf{Proof (by construction of a counterexample):}
To prove this proposition, we construct a counterexample to show that the assumption  adding an edge must lead to a fluctuate or no change in imbalance  is false. We will use a highly symmetric initial network and introduce a  shortcut  that breaks its symmetry to complete the proof.
\begin{enumerate}
    \item \textbf{Initial state: An  ordered  system}
    Consider a \textbf{Ring Graph ($C_8$)} with $N=8$ nodes. This is a highly ordered system where each node has an identical topological position. Intuitively, we expect its imbalance value to be low.
    \begin{itemize}
        \item \textbf{Hop-count distribution}: In $C_8$, the set of shortest path lengths from any node to the others is always {1, 2, 3, 4, 3, 2, 1}. This makes the overall hop-count distribution of the network highly regular, containing only four discrete values {1, 2, 3, 4}, with a very uniform frequency distribution.
    \end{itemize}
    
    % As requested, a note for a figure can be placed here.
    % Suggestion for a two-panel figure: (a) C_8 ring graph, (b) C_8 with the shortcut edge (0,4).

    \item \textbf{Operation: Introducing symmetry breaking}
    We add a  shortcut  edge between node 0 and node 4. This edge is topologically equivalent to the diameter of the ring and greatly disrupts the original rotational symmetry. Nodes 0 and 4 become  special  nodes, while the others do not.

    \item \textbf{Final state: A  functionally more chaotic  system}
    After adding this shortcut, the average path length of the entire network indeed decreases, and the network becomes  more efficient.  However, we are concerned with the \textbf{balance} of connections.
    \begin{itemize}
        \item \textbf{Heterogenization of hop-count distribution}: The new hop-count distribution becomes far more complex and non-uniform. For example, from node 1, the set of hop counts to other nodes becomes  {1, 1, 2, 2, 3, 2, 1} . From node 0, the set is  {1, 1, 2, 1, 2, 1, 1} . The  worldviews  of different nodes begin to differ significantly. The  uniformity  of the global hop-count distribution is destroyed, becoming more heterogeneous and skewed.
    \end{itemize}
    
    \item \textbf{Response of Imbalance}
    We set evaluation parameters, for example, $h_0=3$ and $a=2$.
    \begin{itemize}
        \item In the initial graph $C_8$, the regular hop-count distribution produces a relatively uniform weight distribution ${w_{u,v}}$, which leads to a higher Shannon entropy $H$, and thus a  lower imbalance value.
        \item In the graph with the added shortcut, the highly heterogeneous new hop count distribution produces a more non-uniform weight distribution $\{w _{u,v}\}$ (e.g., many path weights are concentrated towards 1, while a few path weights remain small). This distribution has a higher certainty; thus its Shannon entropy $H $ is lower.
        \item According to the definition of imbalance, a lower entropy $H $ means a higher imbalance value.
    \end{itemize}
\end{enumerate}
The imbalance metric is not a monotonic function of the set of edges. Local topological optimization aimed at improving the overall efficiency of the network may lead to heterogenization of the distribution of the global connection experience by destroying the original topological symmetry. For example, in a highly symmetric ring graph, introducing a  shortcut  edge, although changing the average path length, makes the path length differences between different node pairs larger. This heterogenization of the path distribution will directly lead to a fluctuation in the imbalance value.

\subsection{Relationship with Spectral Graph Properties}
The algebraic connectivity $\lambda_2$ of a graph is a core spectral metric for measuring the robustness of its network structure. This section aims to theoretically establish the intrinsic connection between Imbalance and $\lambda_2$, demonstrating that Imbalance can effectively reflect the structural integrity of the network from the perspective of connection balance.

\textbf{Proposition 2.7}: Let $G=(V,E)$ be a connected graph with $N=|V|$, algebraic connectivity $\lambda_2$, and diameter $\mathrm{diam}(G)$. For any parameters $a>0$ and $h_0>0$, if the choice of parameter $h_0$ satisfies the condition $h_0 > \mathrm{diam}(G)$, then we have: $\lim_{a \to \infty} I(G; a, h_0) = 0$. 
To derive a sufficient condition, we recall the classical upper bound of graph diameter given by \cite{ref24}:

$$diam(G)<\frac{2ln(N-1)}{\lambda_2}$$

where $\lambda_2$ is the algebraic connectivity of the graph. Therefore, any choice of $h_0$ satisfying

$$h_0 > \left\lceil\frac{2\ln(N-1)}{\lambda_2}\right\rceil \implies \lim_{a \to \infty} I(G; a, h_0) = 0$$

\textbf{Proof:} Our proof strategy is to create an ideal condition in which the  connection experience  of all node pairs is identical. To do this, we first set an extremely lenient  QoS threshold $h_0$, making it no less than the diameter of the network, thus ensuring that even the longest path in the network is considered  acceptable.  At the same time, we consider the limit as the sharpness parameter $a \to \infty$, which will make our evaluation criteria extremely  black and white. 

\textbf{Step 1: Setting Parameter Conditions (Hypothesis)}
We set the evaluation parameter $h_0$ to satisfy $h_0 > \mathrm{diam}(G)$, where $\mathrm{diam}(G) := \max_{u,v \in V} d(u,v)$, and consider the limit as the sharpness parameter $a \to \infty$.

\textbf{Step 2: Convergence of weight}
Under the above conditions, since the length of any path $d(u,v)$ does not exceed $h_0$, the exponent of the sigmoid function $a(d(u,v) - h_0)$ must be less than or equal to zero. As the sharpness parameter $a$ approaches infinity, this negative exponent will approach negative infinity, causing the value of the exponential function $e^{(\cdot)}$ to approach 0. Therefore, the connection weights $w(u,v)$ for all pairs of nodes will invariably converge to their maximum value of 1.
\begin{equation} \label{eq:dist_condition}
\forall u \neq v \in V, \quad d(u,v) \le \mathrm{diam}(G) \implies d(u,v) - h_0 < 0
\end{equation}

\begin{equation} \label{eq:limit_a_diff}
\implies \lim_{a \to \infty} a(d(u,v) - h_0) = -\infty
\end{equation}

\begin{equation} \label{eq:limit_wa_one}
\implies \lim_{a \to \infty} w_a(u,v) = \lim_{a \to \infty} \frac{1}{1+e^{a(d(u,v)-h_0)}} = 1
\end{equation}

\textbf{Step 3: Convergence of the probability distribution}
When all connection weights are equal to 1, the total weight of the network $W$ is equal to the total number of pairs of nodes $N(N-1)$. After normalization, the probability share $p(u,v)$ for each pair of nodes therefore becomes completely equal, all equal to $\frac{1}{N(N-1)}$. This is exactly the maximum entropy distribution in information theory.

\begin{align} \label{eq:limit_Wa_and_pa_final}
\lim_{a \to \infty} W_a &:= \lim_{a \to \infty} \sum_{u \neq v} w_a(u,v) \\ % 这里是第一个主要等号的对齐点
&= \sum_{u \neq v} \lim_{a \to \infty} w_a(u,v) \notag \\ % 这里承接上一行，并继续对齐
&= \sum_{u \neq v} 1 = N(N-1) \notag \\[1em] % Wa 的推导完成，添加一些垂直间距，使结构更清晰
\implies \lim_{a \to \infty} p_a(u,v) &= \frac{\lim_{a \to \infty} w_a(u,v)}{\lim_{a \to \infty} W_a} \\ % p_a 的推导开始，对齐到之前等号的位置
&= \frac{1}{N(N-1)} \notag % p_a 的推导完成，继续对齐
\end{align}

\textbf{Step 4: Calculation of entropy and imbalance}
According to the definition of Shannon entropy, a uniform distribution is the distribution with the maximum entropy. Therefore, the entropy of the system $H_a$ will reach its theoretical maximum value $H_{max}$ at this limit. When we substitute this maximum entropy into the definition of normalized entropy $Q$, the resulting value $Q$ is exactly 1. Finally, according to the definition of imbalance $I=1-Q$, we obtain its final value of 0. This mathematically proves that, under the set ideal conditions, the functional fairness of the network reaches a perfect state.

\begin{align} \label{eq:H_and_I_limits}
\lim_{a \to \infty} H_a &= -\sum_{u \neq v} \left(\lim_{a \to \infty} p_a(u,v)\right) \log_2\left(\lim_{a \to \infty} p_a(u,v)\right) \notag \\
&= -\sum_{u \neq v} \frac{1}{N(N-1)} \log_2\left(\frac{1}{N(N-1)}\right) \notag \\
&= \log_2(N(N-1)) := H_{max} \notag \\
\implies \lim_{a \to \infty} I_a &= 1 - \frac{\lim_{a \to \infty} H_a}{H_{max}} \notag \\
&= 1 - \frac{H_{max}}{H_{max}} = 1 - 1 = 0
\end{align}

Q.E.D.

Algebraic connectivity $\lambda_2$ is a key indicator of a network's ability to resist partitioning, and the result of Mohar provides an upper bound for the network diameter controlled by $\lambda_2$. Proposition 2.7 indicates that a more robust structural network is also more likely to exhibit fairness in its functionality, requiring only the setting of a reasonable $h_0$ threshold that can cover its diameter to make its imbalance value approach zero under limit conditions.

While Proposition 2.7 provides a sufficient condition for imbalance vanishing in the asymptotic regime, it relies on carefully tuning the threshold parameter $h_0$ relative to the graph's diameter.From a structural perspective, it is natural to ask whether the imbalance metric can be upper bounded in terms of more intrinsic graph properties, independent of specific parameter values.Noting that the algebraic connectivity $\lambda_2$ characterizes both expansion and average path length, we are led to propose the following conjecture:

\textbf{conjecture 2.8}
\label{conj:spectral_bound}
For any graph $G=(V,E)$ with algebraic connectivity $\lambda_2(G)$, there exists a monotonically non-decreasing function $f: [0, \infty) \to [0, 1)$, which depends only on the Sigmoid parameters $(a, h_0)$ and satisfies $f(0)=0$, such that the Network Imbalance $I(G)$ is upper-bounded by:
$$
I(G) \le 1 - f\bigl(\lambda_2(G)\bigr)
$$

We believe this inequality aligns with the observed trend that stronger spectral connectivity leads to more balanced functional distributions. A rigorous proof of this spectral-to-functional inequality remains an open direction for future research.

\section{Results}

\subsection{Experimental Setup}

To systematically validate the effectiveness and explanatory power of the Network Imbalance (I) metric, we conducted comprehensive simulations across a spectrum of representative network models. Our experimental strategy spans from highly ordered, special graphs (e.g., complete and ring graphs) to the three canonical random graph models (ER, BA, WS) that capture various features of real-world networks. Finally, we apply the metric to a large-scale, real-world network—the Internet's Autonomous System (AS) topology—to assess its practical utility. All simulations were implemented in Python, utilizing the NetworkX, NumPy, and SciPy libraries. The specific parameters for each model, along with the Quality of Service (QoS) profiles used for evaluation, are summarized in Table \ref{tab:sim_params}. The subsequent sections will present and analyze the results of these experiments in detail.

\begin{table*}[htbp]
\centering
\caption{Parameters for Network Models and Simulations}
\label{tab:sim_params}
\begin{tabular}{@{}lllccl@{}}
\toprule
\textbf{Network Model} & \textbf{Parameter} & \textbf{Value} & \textbf{\# of Runs} & \textbf{QoS Profile} & \textbf{Ref. Figure} \\
\midrule
ER Random Graph   & N (Nodes)              & 50               & 20 & (a,$h_0$) varied & Fig. 6  \\
                  & p (Edge Probability)   & [0, 0.4]         & 20 & (a,$h_0$) varied & Fig. 6  \\
\addlinespace
BA Scale-Free     & N (Nodes)              & 50               & 20 & (a,$h_0$) varied & Fig. 7  \\
                  & m (Edges per new node) & {[}1, 10{]}      & 20 & (a,$h_0$) varied & Fig. 7  \\
\addlinespace
WS Small-World    & N (Nodes)              & 50               & 20 & (a,$h_0$) varied & Fig. 8 , 10  \\
                  & k (Initial neighbors)  & 4                & 20 & (a,$h_0$) varied & Fig. 8 , 10  \\
                  & p (Rewiring prob.)     & {[}0, 1{]} (log scale) & 20 & (a,$h_0$) varied & Fig. 8 , 10  \\
\addlinespace
Dumbbell Network  & N (Nodes)              & 50 (2x25 clusters) & 1  & Profile A/B   & Fig. 11  \\
                  & Profile A (Sens.)      & (a=2.0, $h_0$=3)    & 1  & -             & Table II  \\
                  & Profile B (Tol.)       & (a=0.5, $h_0$=6)    & 1  & -             & Table II  \\
\addlinespace
Internet AS       & N (Nodes)              & $\approx$10670   & 1 (snapshot) & (a=1.0, $h_0$=4) & Table III  \\
                  & M (Edges)              & $\approx$22002   & 1 (snapshot) & (a=1.0, $h_0$=4) & Table III  \\
\bottomrule
\end{tabular}
\end{table*}

\subsection{Imbalance on Special Graphs}
In this section, we analyze and calculate its performance on several representative classical graph models. These models, because of their clear structures, are often used as benchmarks for network metric analysis. The final value of the imbalance depends on the parameters $a$ and $h_0$, so our focus is on qualitative analysis and comparison, rather than calculating specific numerical values.

\subsubsection{Complete Graph ($K_N$)}
\begin{figure}[h!]
    \centering
    \includegraphics[width=0.4\linewidth]{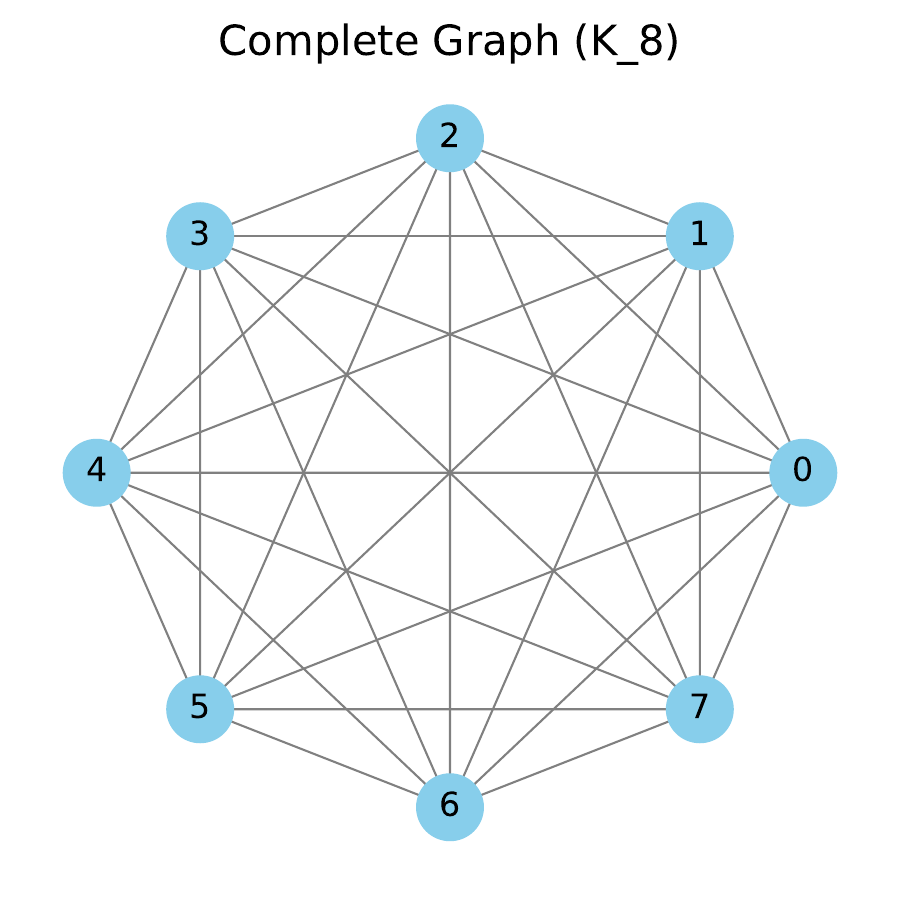} % Renamed for clarity
    \caption{Complete Graph ($K_N$).}
    \label{fig:kn}
\end{figure}

We first examine the behavior of the imbalance metric in the entire graph $K_N$. As the ultimate embodiment of structural symmetry and connection efficiency in network theory, $K_N$ provides an ideal benchmark for our analysis.

In a complete graph, its definition ensures that there is an edge between two distinct nodes $u, v \in V$. This topological property directly leads to the complete degeneracy of its shortest path distribution: the shortest path length between all node pairs is always 1, that is, $h(u,v) = 1, \forall u \neq v$.

According to the definition of the Imbalance metric, when the hop-count distribution of the input network is a single value, regardless of the choice of parameters $a$ and $h_0$, all connection weights $w(u,v)$ obtained through the mapping of the sigmoid function will necessarily be completely equal. The uniformity of the weights further leads to its normalized probability distribution $\{p_{u,v}\}$ becoming a perfect uniform distribution. As strictly proven in \textbf{Proposition 2.4}, a uniform weight distribution will cause the Shannon entropy of the system $H$ to reach its theoretical maximum value $H_{max}$, thus normalizing the entropy $Q=1$.

Therefore, the imbalance value of a complete graph must be zero.
\begin{equation}
I(K_N) \equiv 0
\end{equation}
This result is of significant theoretical importance. Not only is it mathematically consistent with our theoretical derivations, but more importantly, it establishes the zero-point benchmark for the Imbalance metric. This result confirms that our metric can accurately identify $K_N$ - the most ordered in the topological structure and most balanced in function - as the ideal network.

\subsubsection{Path Graph ($P_N$)}
\begin{figure}[h!]
    \centering
    \includegraphics[width=0.4\linewidth]{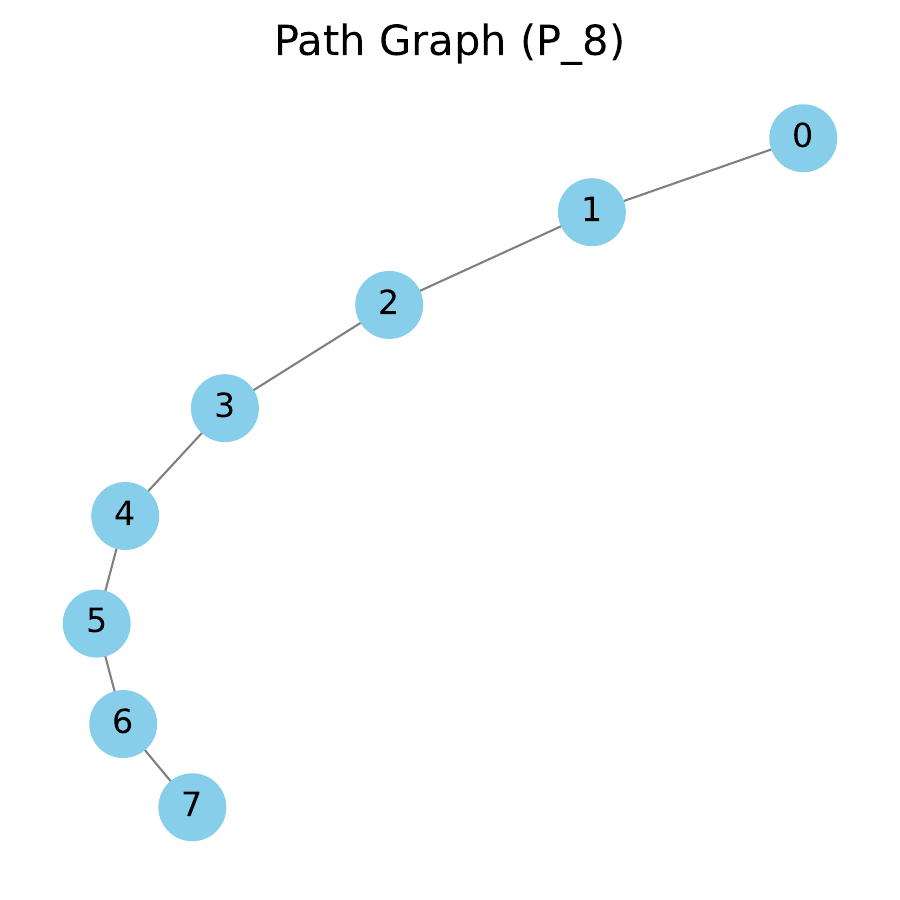} % Renamed for clarity
    \caption{Path Graph ($P_N$).}
    \label{fig:pn}
\end{figure}

In contrast to the extreme symmetry of the complete graph, the path graph $P_N$ represents another topological benchmark, characterized by the sparsity of the connections and the structural asymmetry. In this graph, nodes are arranged linearly, and their topology fundamentally determines the severe imbalance of connection performance.

For any two nodes $v_i$ and $v_j$ in $P_N$, their shortest path is unique and its length is $h(v_i, v_j) = |i-j|$. This property leads to a completely different multi-pair hop-count distribution from that of $K_N$: this distribution is no longer a single point, but spans the entire integer range from 1 to $N-1$. More importantly, the distribution is highly skewed, with the frequency of short paths much higher than that of long paths.

This inherent and broad heterogeneity in the hop-count distribution is the fundamental reason for the high imbalance value of the path graph. When this wide spectrum of hop counts is mapped through the monotonic Sigmoid function, it inevitably produces an equally wide and nonuniform spectrum of connection weights $\{w(u,v)\}$. This weight structure, far from uniform, greatly suppresses the overall Shannon entropy $H$ of the system, leading to a small value of the normalized entropy $Q$.

Therefore, the imbalance value of the path graph is necessarily significantly greater than zero and, under the appropriate parameters, approaches its theoretical upper limit as $N$ increases. This higher imbalance value is a quantitative description of the inherent and severe  positional inequality  within this topology. It accurately captures that central nodes in the network have an unparalleled connection advantage over peripheral nodes at both ends, leading to a severe imbalance in end-to-end accessibility on a global scale.

\subsubsection{Ring Graph ($C_N$)}
\begin{figure}[h!]
    \centering
    \includegraphics[width=0.4\linewidth]{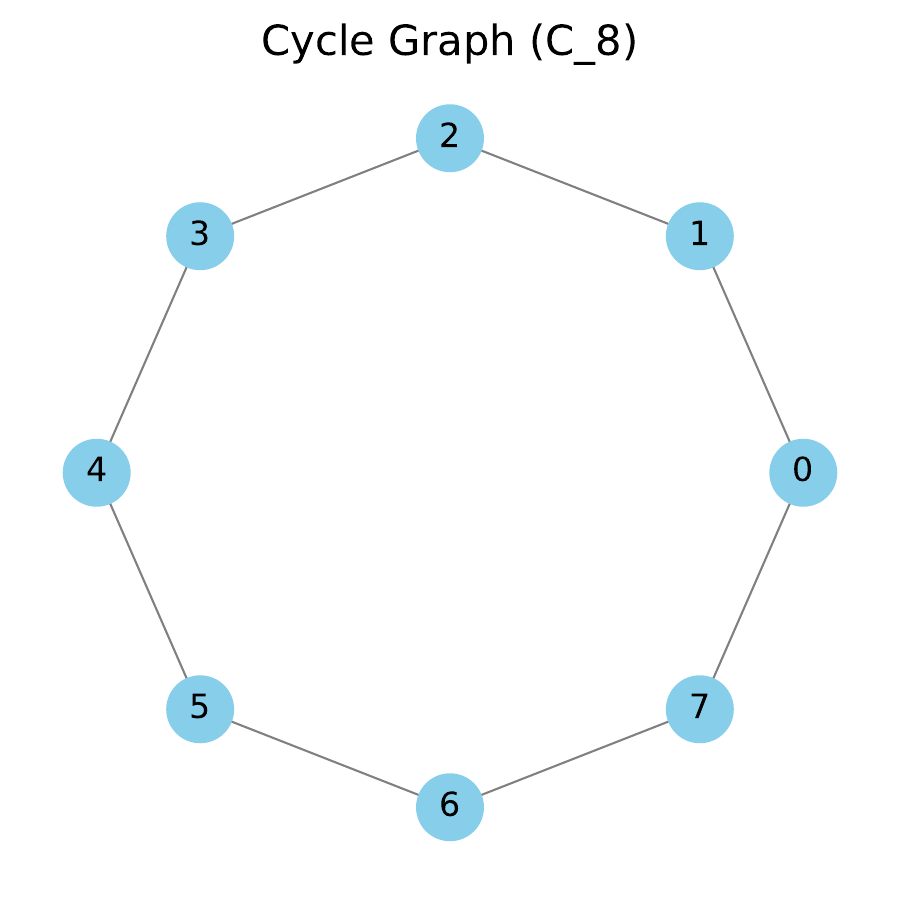} % Renamed for clarity
    \caption{Ring Graph ($C_N$).}
    \label{fig:cn}
\end{figure}

The ring graph $C_N$ is a minimal topological improvement over the path graph $P_N$ that connects the two endpoints of the path graph. However, this simple closure operation brings about a significant improvement in functional symmetry. It forms a 2-regular graph and is \textbf{vertex-transitive}, meaning that the network looks identical from the  perspective  of any node, which eradicates the severe positional inequality caused by the  end effects  in the path graph.

Compared to the path graph, the connection structure of the ring graph is more optimized. The distance between any two nodes is compressed to $h(v_i, v_j) = \min(|i-j|, N-|i-j|)$, and therefore the diameter of the network is halved to $\lfloor N/2 \rfloor$. This structural improvement makes the all-pairs shortest-path (APSP) distribution have a  narrower range and a more symmetric and uniform shape . A more compact and balanced path distribution, after being transformed by the Sigmoid function, will inevitably produce a connection weight distribution with smaller variance and greater uniformity. Therefore, the system's normalized Shannon entropy  Q  value will be significantly higher  than that of a path graph with the same number of nodes, ultimately leading to its  Imbalance  value being  much lower than that of the path graph : $I(C_N) \ll I(P_N)$.

However, it should be noted that unless it is the trivial case of $N=3$ (where $C_3$ is isomorphic to $K_3$), the APSP distribution of the ring graph still contains multiple discrete values and has not reached the single-point degenerate state of the complete graph $K_N$. This means that its functional fairness has not yet reached perfection and the normalized entropy  Q  cannot reach its theoretical maximum of 1, so its  Imbalance  value must be  strictly greater than zero .

The precise capture of this subtle difference by the  Imbalance  metric highlights its value as a high-precision  ruler : it can not only distinguish between the extremes of order and disorder but also differentiate between different levels of  order - from the high imbalance of the path graph, to the approximate balance of the ring graph, to the perfect balance of the complete graph. $C_N$ thus becomes a key benchmark for verifying the resolving power of this metric.

\subsubsection{Star Graph ($S_N$)}
\begin{figure}[h!]
    \centering
    \includegraphics[width=0.4\linewidth]{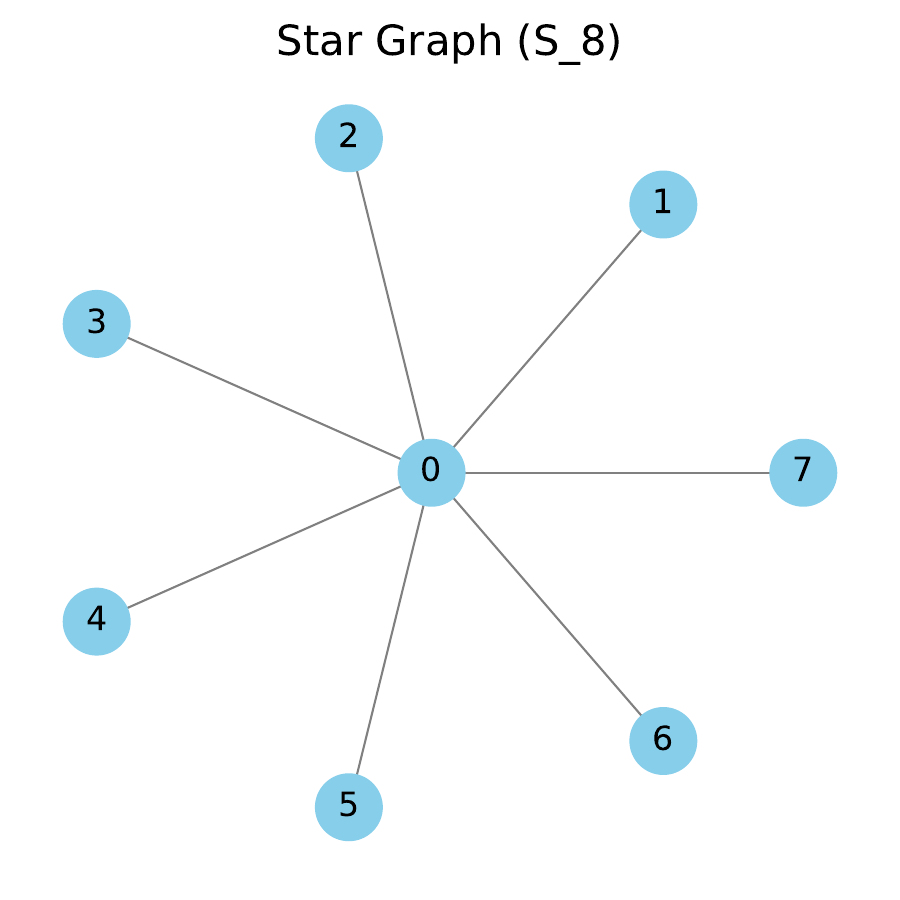} % Renamed for clarity
    \caption{Star Graph ($S_N$).}
    \label{fig:sn}
\end{figure}
The star graph $S_N$ provides an ideal model for studying extremely centralized topologies. It consists of a central node connected to $N-1$ peripheral nodes (leaves), which are only connected to the center. This unique  core-periphery  architecture results in a very simple bimodal distribution of all pairs  shortest paths (APSP), with path lengths having only two possible values: 1 (between the center and peripheral nodes) and 2 (between any two peripheral nodes).

Although the variety of path lengths is extremely small, their frequency distribution has a significant intrinsic asymmetry. The total number of paths of length 1 is $2(N-1)$, while the total number of paths of length 2 is $(N-1)(N-2)$. For any network with $N>3$, the latter number is much higher than the former. When this bimodal but frequency-imbalanced distribution is mapped through the Sigmoid function, it inevitably produces a similarly highly skewed weight probability distribution $\{p_{u,v}\}$. Even if parameters are chosen (e.g. $h_0=1.5$) to perfectly separate these two types of paths, the event probabilities corresponding to the two types of weights will be highly imbalanced due to the huge difference in their cardinalities. According to information theory principles, this highly skewed probability distribution will cause the system's normalized Shannon entropy $Q$ to deviate significantly from 1.

Therefore, star graphs typically exhibit a significantly non-zero  Imbalance  value ($I=1-Q$). It is important to emphasize that the  imbalance  revealed by the  I  metric here has a different origin from the  positional disadvantage  in the path graph. Instead, it arises from the extreme importance (criticality) and structural dominance of the central node. This central node is a mandatory point for all communication paths between the vast majority of node pairs, becoming the network's structural bottleneck. The  Imbalance  metric accurately captures that this extreme centralization is a severe deviation from functional fairness, thereby identifying this topology as a functionally highly imbalanced system.

\subsection{Imbalance Behavior in Random Graph Models}

This chapter places the imbalance metric in the  controlled experimental environments  defined by the three most fundamental random graph models, aiming to systematically reveal its profound ability as a probe of network structure and function. We will no longer just passively observe, but actively pose and answer three core questions:
\begin{enumerate}
    \item In the Erds-Rényi (ER) model, can the I metric sensitively capture the percolation transition of the network from fragmented to connected?
    \item In the Barabási–Albert (BA) model, how does the I metric interpret the paradox between functional efficiency and structural heterogeneity caused by the  rich-get-richer  mechanism?
    \item In the Watts–Strogatz (WS) model, can the I metric accurately describe the complete, nonmonotonic evolutionary path of a network from regular order, through the chaos of symmetry breaking, to a new order of random efficiency?
\end{enumerate}
Through these three major  benchmarks,  this chapter aims to prove that imbalance is not just a static metric, but a sensitive  sensor  capable of dynamically capturing the evolution of macroscopic network properties.

\subsubsection{Erdős–Rényi (ER) Model}
\begin{figure}[h!]
 \includegraphics[width=\columnwidth]{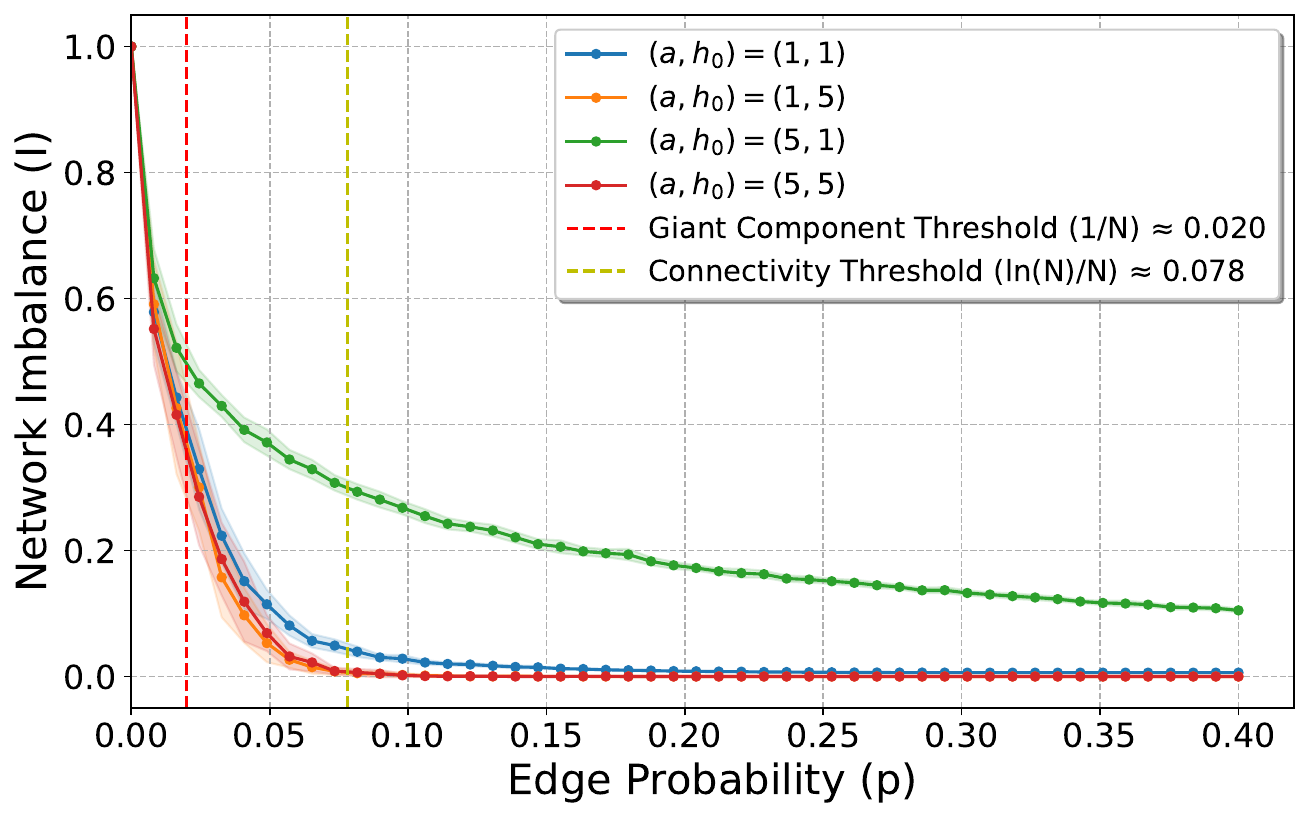}
 \caption{\label{fig:er} Robust simulation results in ER random graphs (N=50, 20 independent experiments). The solid line represents the average trend of the Imbalance value, and the shaded area represents a one-standard-deviation range of fluctuation.}
\end{figure}
The Erdős–Rényi (ER) random graph provides the purest theoretical model for studying how randomness generates global order. Its most famous property is the Percolation Transition, where the network undergoes a sudden shift from a collection of fragmented subgraphs to the emergence of a macroscopic Giant Component as the connection probability  p  crosses the critical point $p_c \approx 1/N$. The core task of this section is to test whether our imbalance metric, as a functional measure, can sensitively capture this purely structural phase transition.

The behavior of the imbalance metric confirms the percolation transition process. In the subcritical region, the I value stabilizes near its maximum of 1. Near the critical point $p_c \approx 1/N$, the I value undergoes a characteristic sharp drop. As the network enters the supercritical region, the I value smoothly converges towards 0. The simulation results show that the average trend of the entire process is very stable, with minimal random fluctuations (shaded area), demonstrating the reliability of the I metric in capturing this macroscopic phase-transition phenomenon.

\subsubsection{Barabási–Albert (BA) Model}
\begin{figure}[htbp]
 \includegraphics[width=\columnwidth]{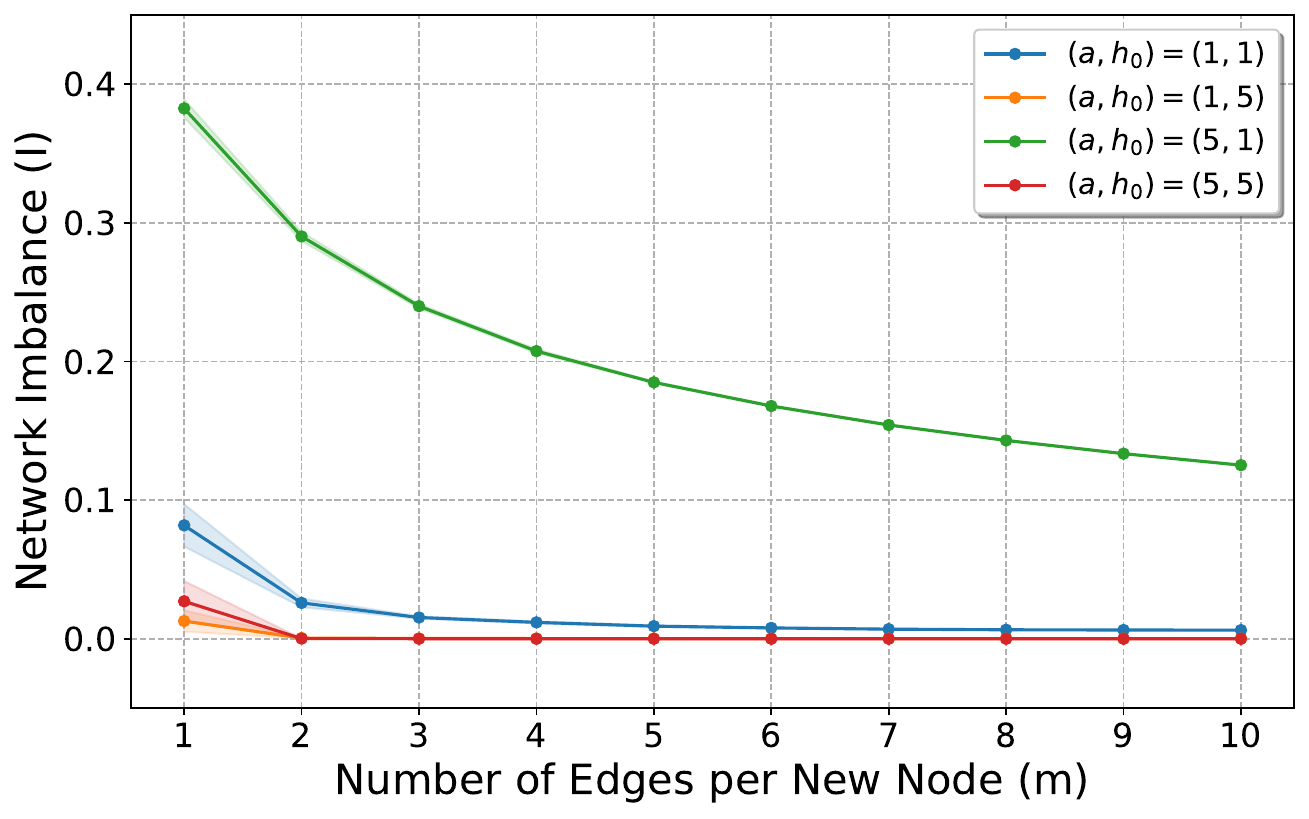}
 \caption{\label{fig:ba} Robust simulation results in BA scale-free networks (N=50, 20 independent experiments).}
\end{figure}
The simulation results and theoretical analysis jointly reveal a clear conclusion that depends on the evaluation perspective: under most  reasonable  evaluation criteria, BA networks exhibit extremely low functional imbalance; however, when service requirements are extremely stringent, they instead show high functional unfairness.

The low I value in most cases is due to extreme structural inequality, leading to a high degree of functional equality through ultimate efficiency. The existence of a few  hub  nodes, like a  highway  system running through the network, greatly compresses the communication paths across the entire network, causing the vast majority of shortest path lengths to be  homogenized  within a very small range. When our evaluation  ruler  $h_0$ is set higher than this range, the vast majority of connection experiences are considered  satisfactory,  and their weight w values are highly homogenized, thus leading to an I value approaching zero.

However, when the quality of service threshold is set to be extremely stringent (e.g. $h_0=1$), the I value will increase significantly. This is because, although hub nodes improve efficiency, they do so at the cost of sacrificing direct connections. From a perspective that does not tolerate any transit, the functional experience of a BA network is precisely unfair. The Imbalance metric sensitively captures this QoS-dependent  role reversal.  Therefore, the analysis of the BA model using the I metric provides the most profound illustration of  structure-function decoupling. 

\subsubsection{Watts–Strogatz (WS) Model}
\begin{figure}[h!]
 \includegraphics[width=\columnwidth]{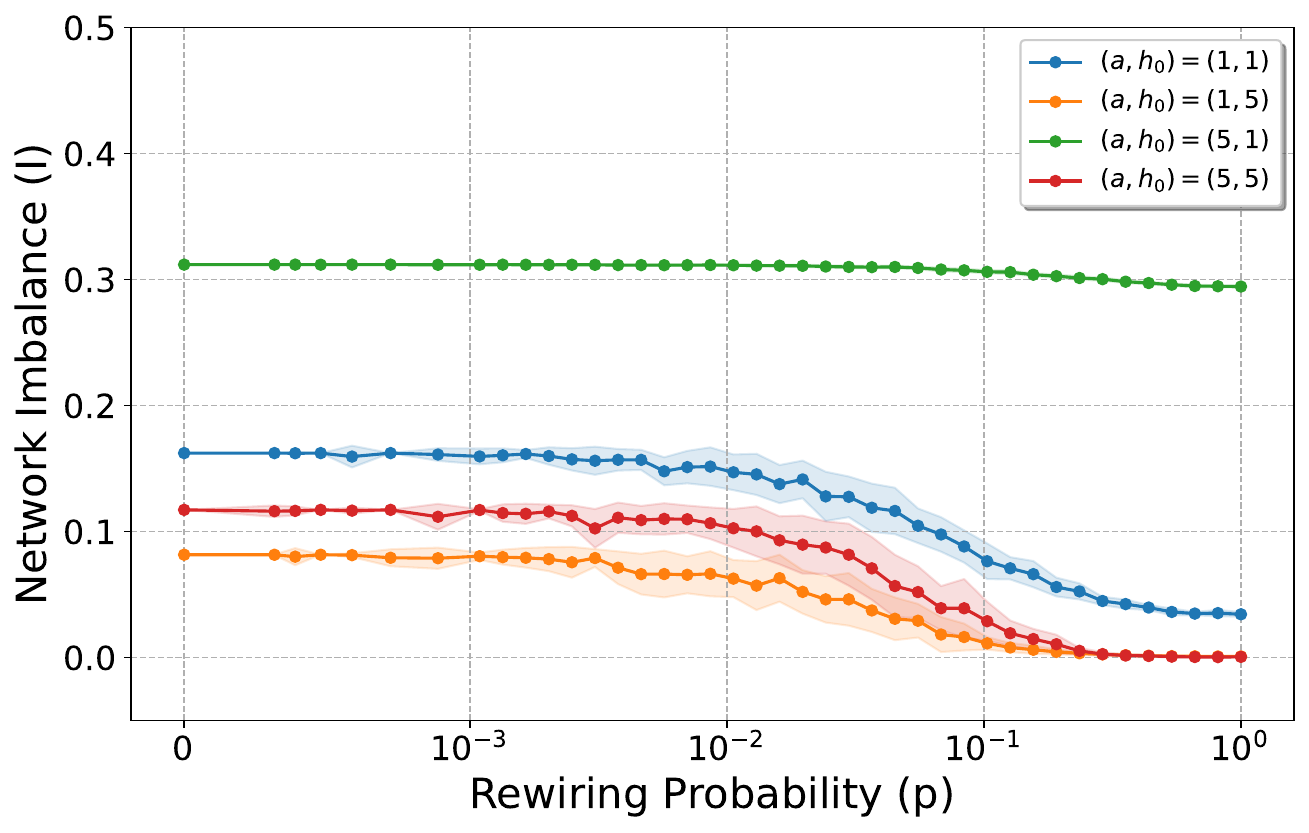}
 \caption{\label{fig:ws} Robust simulation results in WS small-world networks (N=50, k=4, 20 independent experiments).}
\end{figure}

The Watts-Strogatz (WS) model provides an ideal experimental platform for testing how the interplay between efficiency improvement and symmetry breaking affects functional fairness. 

As shown in Fig.~8, our simulation results clearly show that for all combinations of parameters, the average trend of the imbalance value exhibits a strong and predominantly decreasing behavior on a macroscopic scale. It is worth noting that the shaded area in the figure represents the statistical fluctuations of individual simulation runs. This phenomenon reveals behaviors on two distinct levels:
\begin{enumerate}
    \item \textbf{Individual Realization Level:} For any single, specific network evolution (i.e., the process of random rewiring), the $I$ value can certainly exhibit local fluctuations due to the stochastic nature of symmetry breaking. This is entirely consistent with the non-monotonicity principle revealed by our counterexample in Proposition~2.6.
    \item \textbf{Statistical Level:} However, when averaged over many experiments, the ``global efficiency-gain'' effect introduced by random shortcuts systematically overwhelms the ``local symmetry-breaking'' effect. While minor local non-monotonicities can be observed even in the average curves for some parameters within the transition region, this does not alter the overall conclusion that the efficiency gain is the dominant factor statistically. The introduction of shortcuts universally and substantially shortens network paths, causing the overall path distribution to become more compact and leading to a globally decreasing trend for the average $I$ value.
\end{enumerate}
Therefore, these simulation results do not invalidate the symmetry-breaking principle but rather indicate that, for a network of this size (e.g., $N=50, k=4$), the efficiency gain is the dominant factor that shapes the overall statistical trend.

\section{Discussion}
\subsection{Relationship with the Degree Distribution}
\begin{figure}[tbp]
 \includegraphics[width=\columnwidth]{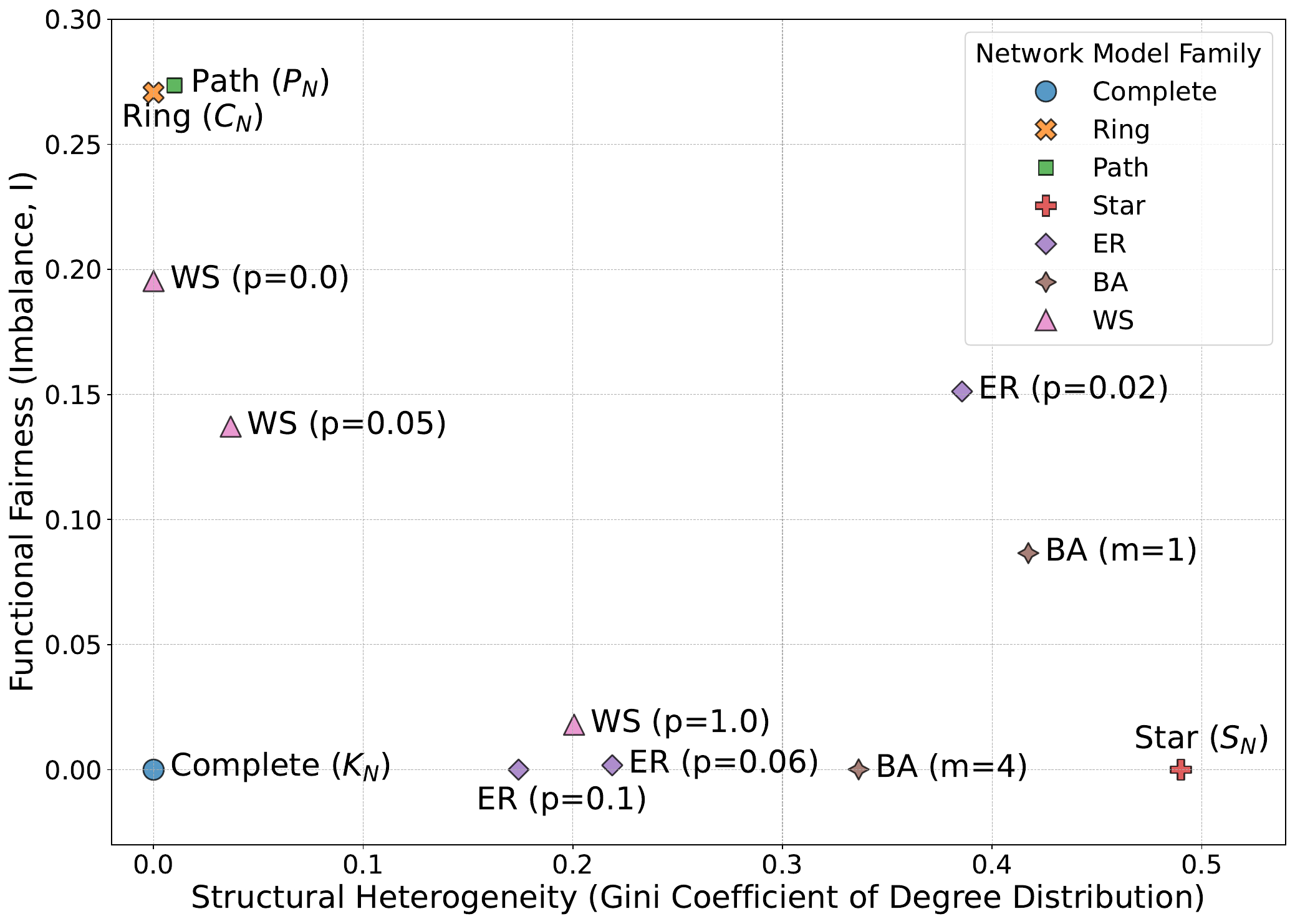} % Renamed for clarity
 \caption{\label{fig:zoo}  Decoupling of structural heterogeneity and functional fairness across a "zoo" of network models(a=3,$h_0=4$).Structural heterogeneity is quantified by the Gini Coefficient of the degree distribution, where a higher value indicates greater inequality.}
\end{figure}

Unlike the hop-count distribution, the Imbalance metric does not have a direct functional relationship with the network degree distribution $P(k)$, but exhibits a more profound and complex coupling. This section aims to argue that there is a significant decoupling between a network's structural heterogeneity (determined by the shape of $P(k)$) and its functional fairness (measured by the $I$-value). This finding highlights the unique value of the imbalance metric in providing a novel functional perspective that goes beyond traditional structural measures.To visually demonstrate this decoupling, we simulated a "zoo" of network models and mapped their structural and functional properties onto a single landscape, as shown in Fig. \ref{fig:zoo}.

A low imbalance state in a network, that is, a high functional equilibrium, can be achieved through two distinctly different paths. The first path is through structural symmetry. In regular graphs (such as ring or complete graphs shown on the left side of Fig. \ref{fig:zoo}), nodes are topologically equivalent, their degree distribution is extremely concentrated, and the structure is highly uniform. This inherent symmetry naturally leads to a balanced path length distribution, thus resulting in a very low I value.

The second path is through extreme efficiency. In scale-free networks with a power-law degree distribution (such as the BA model in Fig. \ref{fig:zoo}), the network is structurally highly heterogeneous, with a few  hub  nodes having extremely high degrees. However, these hub nodes act as global  shortcuts,  greatly compressing the distances between all node pairs in the network, causing the vast majority of shortest path lengths to be  homogenized  within a very small range. This functional efficiency, driven by structural inequality, also results in a very low $I$ value.

Therefore, we arrive at a core conclusion: structural homogeneity (as in regular graphs) and structural heterogeneity (as in scale-free networks), two diametrically opposed organizational forms, can both lead to the same result of functional homogeneity (a low I value). The imbalance metric can accurately capture these two different underlying mechanisms, the first stemming from symmetry and the second from efficiency. This reveals the limitations of inferring a network's functional fairness solely from structural indicators like the degree distribution.

\subsection{Comparison with Classical Metrics}
A natural question is: what is the uniqueness and necessity of the Imbalance metric compared to traditional metrics such as path length variance (structural inequality), average path length (efficiency), and the classic Jain's Fairness Index (fairness)? To answer this question, we designed a comprehensive comparative experiment, observing the behavior of these four types of metrics simultaneously during the evolution of the Watts-Strogatz small-world model.

\begin{figure}[tbp]
 \includegraphics[width=\columnwidth]{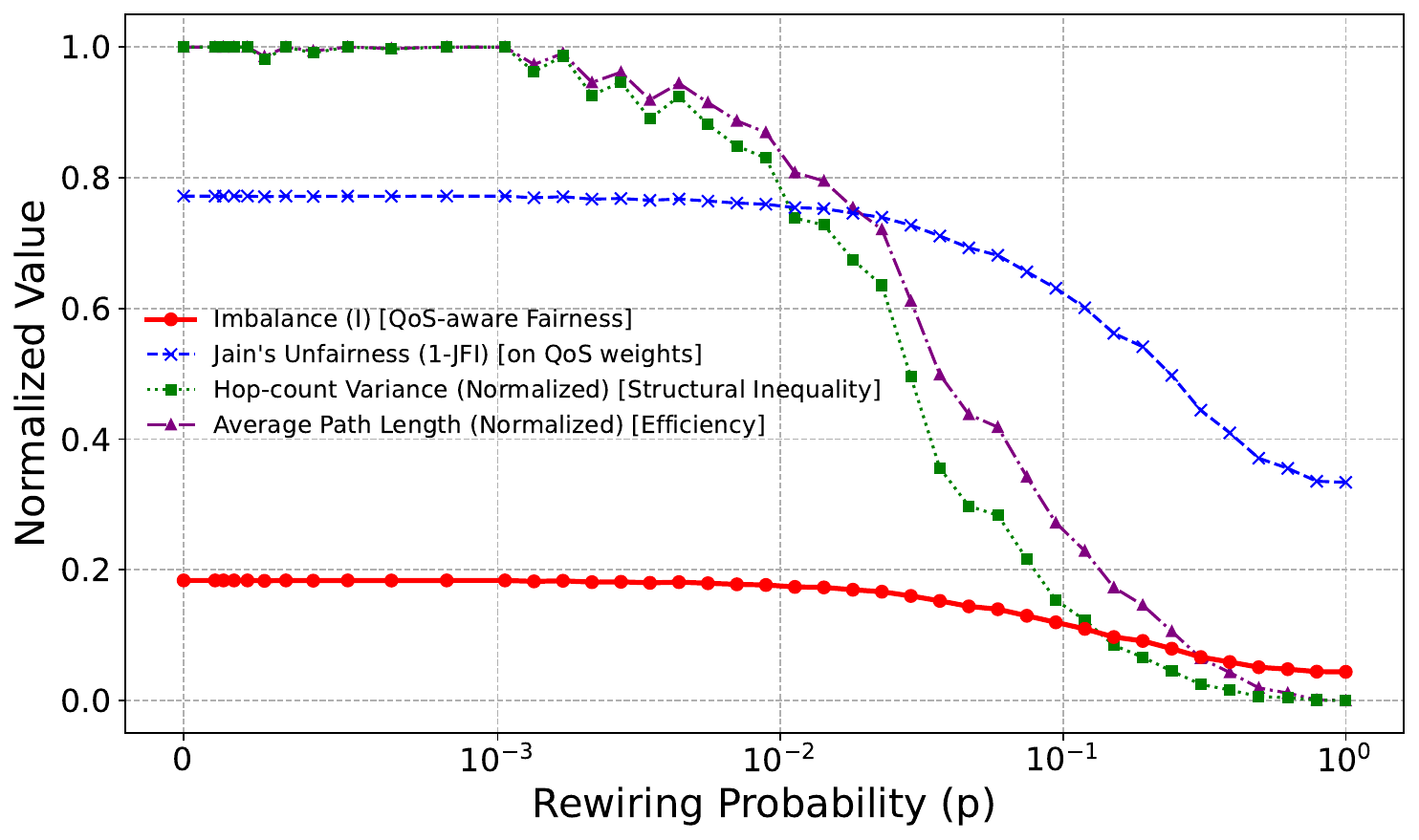} % Renamed for clarity
 \caption{\label{fig:comparison} Comparison of the behavior of Imbalance and three classical network metrics during the evolution of the WS model (N=50, k=4). Jain's Unfairness (1-JFI) is calculated on the same set of QoS weights ${w(u,v)}$ for direct comparison.}
\end{figure}

As shown in Fig. \ref{fig:comparison}, this comparative experiment clearly reveals four core features of the Imbalance metric, thus demonstrating its irreplaceable theoretical value:
\begin{enumerate}
    \item The curve of Imbalance (I) is qualitatively highly consistent with that of Jain s Unfairness (1 - JFI) applied to the same set of QoS weights, jointly confirming that as the network evolves towards randomization, its functional fairness improves unidirectionally. This demonstrates the basic validity of I as an effective fairness metric. However, the quantitative and morphological differences between the two highlight the unique theoretical value of the I metric. The I metric is based on information entropy and measures the overall uniformity of a distribution, whereas JFI is based on the second moment and is more sensitive to the variance of the distribution.
    \item The shape of the I value curve is completely different from that of the pure  hop-count variance  curve. When the p-value is small, the network exhibits extremely high path length variance due to its large diameter (high structural inequality), but its imbalance value is at a relatively moderate level. This decisively proves that Imbalance is not a simple substitute for path length variance. By introducing the QoS  lens  of ($a, h_0$), the I metric can distinguish between a  meaningful  inequality and a  harmless  inequality, thus achieving a  context-aware  evaluation.
    \item The sharp decrease in average path length (representing efficiency) only indicates that the network is becoming  faster,  but it cannot tell us whether the  dividends  of this speed-up are fairly distributed. The gradual decrease in the I value shows that the improvement in functional fairness is a more gradual and robust process than the improvement in efficiency.
    \item The core advantage of the entire theoretical framework of Imbalance lies in its endogenous nature. Through the tunable parameters ($a, h_0$), it incorporates QoS requirements into the fairness evaluation. What it measures is not the fairness of the raw network topology, but the functional fairness filtered through a specific functional  ruler. 
\end{enumerate}
In summary, this comparative experiment eloquently demonstrates that the Imbalance metric successfully carves out a new, indispensable fourth dimension centered on QoS-aware  functional fairness  within the  efficiency-structure-fairness  triangular relationship of network evaluation.

\subsection{Case Study: Topology Optimization based on Imbalance}
To validate the application potential of the Imbalance metric in network engineering practice, this section designs and executes a topology optimization case study. Its purpose is to demonstrate that Imbalance can not only effectively quantify the adaptability of a topology to different  QoS requirements, but can also serve as an objective function to guide the optimization of network structures.

\subsubsection{Experimental Design}
\begin{figure}[tbp]
    \centering
    \begin{subfigure}[b]{0.8\columnwidth}
        \centering
        \includegraphics[width=0.7\textwidth]{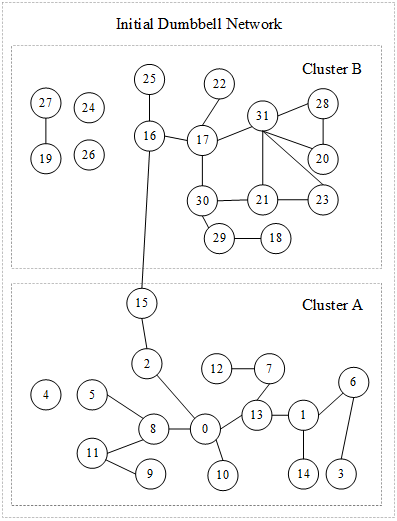}
        \caption{Initial topology $G_0$}
        \label{fig:dumbbell_a}
    \end{subfigure}
    \vfill
    \begin{subfigure}[b]{0.48\columnwidth}
        \centering
        \includegraphics[width=\textwidth]{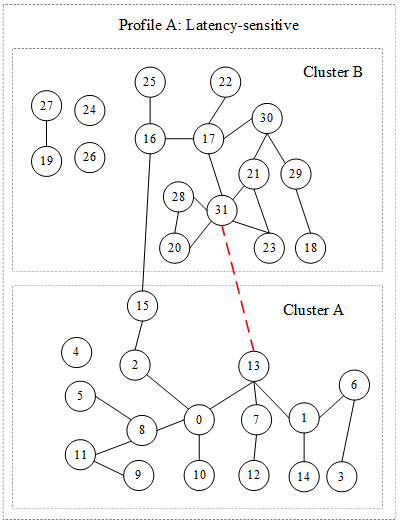}
        \caption{Optimized topology for Profile A}
        \label{fig:dumbbell_b}
    \end{subfigure}
    \hfill
    \begin{subfigure}[b]{0.48\columnwidth}
        \centering
        \includegraphics[width=\textwidth]{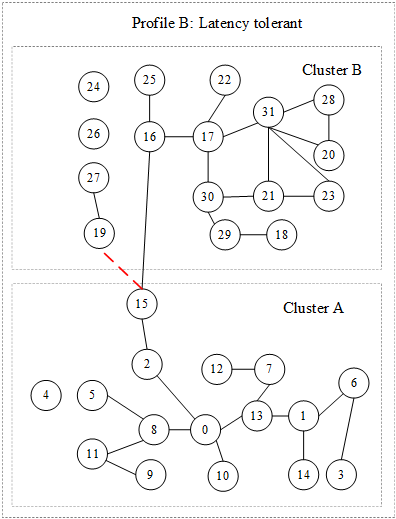}
        \caption{Optimized topology for Profile B}
        \label{fig:dumbbell_c}
    \end{subfigure}
    \caption{Topology optimization process of the dumbbell network.}
    \label{fig:dumbbell}
\end{figure}

\textbf{1. Network Model} We use a typical Dumbbell Network $G_0$ with a clear topological bottleneck as the initial topology. This network consists of two submodules, Cluster A and Cluster B, each with 25 nodes. The two clusters are connected by a single bridge edge, forming a connected graph with $N=50$ nodes, as shown in Fig. \ref{fig:dumbbell_a}. This model is a simplified abstraction of data center or cross-site network connection scenarios, and its structural bottleneck is easy to analyze.

\textbf{2.  QoS Profiles} To simulate heterogeneous traffic, this study defines two service profiles with significantly different QoS requirements:
\begin{itemize}
    \item \textbf{Profile A: Latency-sensitive}: This profile represents applications with strict network latency requirements, such as real-time voice (VoIP) or online gaming. Its imbalance evaluation parameters are set to an ideal hop count threshold $h_0 = 3$ and a steepness parameter $a = 2.0$, to impose a significant weight penalty on paths longer than 3 hops.
    \item \textbf{Profile B: Latency tolerant}: This profile represents applications that are not sensitive to latency, such as large file transfers or nonreal-time data backups. Its evaluation parameters are set to an ideal hop count threshold $h_0 = 6$ and a steepness parameter $a = 0.5$, indicating a higher tolerance for longer paths.
\end{itemize}

\textbf{3. Optimization goal} The optimization goal of this case study is to maximize the adaptability of the network topology to the most latency-sensitive Profile A, under the constraint of being allowed to add one redundant link. This goal is mathematically equivalent to minimizing the imbalance value for profile A, that is, $\min I_A(G )$.

\subsubsection{Initial Topology Evaluation}
The evaluation of the initial topology $G_0$ requires first to calculate the shortest paths between all its pairs of nodes (APSP). The analysis shows that the distribution of the path length of $G_0$ is extremely uneven, as a large number of communication paths between clusters must pass through the single bridge edge, leading to significantly increased path lengths.

Based on this path distribution, the calculated initial imbalance values are shown in Table \ref{tab:opto_results}. The results show that $I_A(G_0) = 0.179$, a value far from the ideal zero, indicates serious connection balance issues for latency-sensitive services. In contrast, $I_B(G_0) = 0.064$, which is at a relatively low level, is consistent with Profile B's higher tolerance for long paths. The evaluation results quantitatively confirm the negative impact of the topology's structural bottleneck on high-performance services.

\subsubsection{Imbalance-driven Topology Optimization}
To improve network topology, this study employs a Greedy Edge Addition algorithm with imbalance minimization as its objective. The algorithm is defined as follows:
\begin{itemize}
    \item \textbf{Objective Function}: $\min_{e  \in E_{\text{cand}}} I_A(G_0 \cup {e })$
    \item \textbf{Input}: Initial topology $G_0$, candidate edge set $E_{\text{cand}}$ (the set of all nonexistent edges in $G_0$).
    \item \textbf{Output}: Optimal edge to add, $e^{*}$.
    \item \textbf{Process}: The algorithm iterates through each candidate edge $e _i$ in $E_{\text{cand}}$. In each iteration, it constructs a temporary graph $G _i = G_0 \cup {e _i}$ and calculates its $I_A(G _i)$ value. Finally, it selects the candidate edge that minimizes the $I_A$ value as $e^*$.
\end{itemize}
After execution, the algorithm identifies the optimal edge to add, $e^*$, as an inter-cluster link connecting non-bridge nodes in Cluster A and Cluster B. The optimized network topology $G_1 = G_0 \cup {e^*}$ provides a crucial redundant path for inter-cluster communication.

% NOTE: As per your request, the following three images are placed within a single figure environment using subfigure.

\subsubsection{Analysis of Optimization Results}
The optimized topology $G_1$ was re-evaluated for imbalance, with the results shown in Table \ref{tab:opto_results}.
\begin{table}[h!]
\caption{\label{tab:opto_results}Comparison of Imbalance values for the two service profiles before and after optimization}
\begin{ruledtabular}
\begin{tabular}{lcc}
Topology State & $I_A (h_0=3, a=2.0)$ & $I_B (h_0=6, a=0.5)$ \\
\colrule
Initial Topology $G_0$    & 0.179 & 0.064 \\
Optimized Topology $G_1$  & \textbf{0.133} & 0.046 \\
\end{tabular}
\end{ruledtabular}
\end{table}

The optimization results show that the Imbalance value for Profile A decreased from 0.179 to 0.133. This significant reduction demonstrates that the optimization algorithm, guided by Imbalance, successfully located and mitigated the structural bottleneck of the network. The newly added redundant link significantly shortened the previously long inter-cluster paths, greatly improving the balance of the overall path distribution.

Furthermore, the Imbalance value for Profile B also improved, decreasing from 0.064 to 0.046. This phenomenon indicates that topology optimization targeting specific strict QoS requirements can often have a universally positive impact on the overall health of the network connection. This case study confirms that the imbalance metric is not only an effective tool for assessing network state but can also serve as an operational objective function to guide automated network design and optimization tailored to specific service requirements, successfully applying the abstract principle of  connection balance  to engineering practice.

\subsection{Analysis of Real-World Networks}
\subsubsection{Motivation and Network Selection}
To test the validity of the  Imbalance  metric beyond theoretical models and to explore its explanatory power in real complex systems, this section analyzes a critical and highly significant infrastructure network: the global Internet's Autonomous System (AS) topology. The Internet AS topology depicts the macroscopic map of global network connectivity, where nodes represent independent networks (that is, ASs, such as Google, Comcast, or a university network), and edges represent their peering or transit protocol relationships.

\subsubsection{Data and Methodology}
This study uses a publicly available AS relationship dataset (e.g., from the CAIDA project) to construct the network topology. To maintain consistency with the theoretical framework of this paper, we treat the AS network as an undirected, unweighted graph $G_{AS}$, where nodes are ASs and an edge exists between two ASs if there is at least one peering or customer-provider relationship between them.

We selected an AS topology snapshot that contains $N \approx 10670$ nodes and $M \approx 22002$ edges. The analysis procedure is as follows:
\begin{enumerate}
    \item Calculate the shortest path count (APSP) of all pairs $h(u,v)$ for this network.
    \item Set a representative set of evaluation parameters for the quality of the wide area network connection. Considering the  small-world  nature of the Internet, we set the ideal hop count threshold at $h_0 = 4$ and the steepness parameter to $a=1.0$, representing a general QoS requirement that is relatively satisfied with accessibility within 4 hops and has a gradually decreasing tolerance for longer paths.
    \item Based on these parameters, calculate the  Imbalance  value $I(G_{AS})$ of the AS network.
\end{enumerate}

\subsubsection{Results and Analysis}
The calculated  Imbalance  value of the Internet AS topology is compared with that of an ER random graph of the same scale (with connection probability $p$ set near its connectivity threshold) and a small world model WS (in its  most chaotic  state at the  Imbalance  peak). The results are shown in the table below.

\begin{table}[h!]
\caption{\label{tab:real_world}Comparison of Imbalance values between the Internet AS topology and theoretical models ($h_0=4, a=1.0$).}
\begin{ruledtabular}
\begin{tabular}{lccl}
Network Model & Node Count (N) & $I$ Value  \\
\colrule
ER Random Graph  & $\sim$10670 & $\sim$0.0255  \\
WS Small-World  & $\sim$10670 & $\sim$0.3146 \\
\textbf{Internet AS} & \textbf{$\sim$10670} & \textbf{$\sim$0.0034}  \\
\end{tabular}
\end{ruledtabular}
\end{table}

The results clearly show that the Internet AS topology exhibits an \textbf{extremely low degree of functional imbalance} ($I \approx 0.0034$). This value is not only far lower than that of theoretical models in a  chaotic  state but is even close to the level of an ideal, structurally perfect symmetric network.

\subsubsection{Discussion and Interpretation}
This remarkably low imbalance value provides strong empirical support for the  structure-function decoupling  argument proposed in Section 4.1. The Internet AS topology is structurally highly unequal, with a Gini coefficient for its degree distribution being high, and a few Tier-1 backbone network operators occupying an absolutely dominant position in network connectivity. However, the Imbalance metric reveals that this structural inequality is precisely the source of its high degree of functional fairness. This is perfectly consistent with the phenomenon that we observed in the BA model:
\begin{itemize}
    \item A few  hub  AS nodes (Tier-1 ISPs) form an  information superhighway  that spans the globe, greatly compressing the routing hops between any two ASs.
    \item This causes the entire network s APSP distribution to be  homogenized  within a very small diameter range (typically 3-5 hops).
    \item Therefore, from a reasonable QoS perspective (e.g., $h_0=4$), the vast majority of end-to-end  connection experiences  are of high quality and highly similar, their weights $w(u,v)$ tend to be consistent, thus resulting in an extremely low  Imbalance  value.
\end{itemize}
It can be said that the  Imbalance  metric accurately quantifies the core philosophy of Internet design: \textbf{to provide a universally efficient and fair connection service to all users by building a hierarchical and structurally unequal backbone network.}

It is worth highlighting that if we were to adopt an extremely stringent evaluation standard (e.g., $h_0=1$, requiring all ASs to be directly connected), the  I  value of the AS network would rise sharply. In summary, empirical analysis of the Internet AS topology successfully links the theoretical insights of the  Imbalance  metric with the organizational principles of complex systems of real-world use, verifying the effectiveness and profound explanatory power of this framework as an advanced network evaluation tool.

\section{Conclusion}
The core contribution of this study is the revelation of a duality in network science: there is more than one path to  functional fairness  (a low Imbalance value). A system can achieve functional equilibrium either through structural symmetry (as in a complete graph) or through connection efficiency driven by structural inequality (as in a scale-free network). This  structure-function decoupling  phenomenon is the core theoretical insight provided by the Imbalance metric, offering a new perspective for re-understanding the  order  and  health  of networks. Furthermore, through simulations on the WS small-world model, we analyzed the impact of  symmetry breaking  on functional fairness. The results indicate that for a global metric like I, the efficiency gains from shortcuts are the dominant factor, systematically and unidirectionally optimizing the network's functional fairness. Finally, through validation in application scenarios such as topology optimization, robustness evaluation, and critical link identification, the results show that the I metric is not only an effective tool for assessing network state but can also serve as an operational diagnostic tool and design objective function, applying the principle of  connection balance  to engineering practice. We believe that this metric and its subsequent research can provide valuable references for fields such as network science and systems engineering.

\subsection{Decoupling of Structure and Function and a New View of Network  Order }
The most central theoretical contribution of this study is the revelation of the profound decoupling between network  structural heterogeneity  and  functional homogeneity.  Through analysis on classical graph models, we have shown that a low imbalance value, that is, a high degree of functional equilibrium, can be achieved through two diametrically opposed paths: one, through perfect structural symmetry (as in a complete graph); the other, through extreme connection efficiency driven by  hub  nodes (as in a scale-free network). The profoundness of the  Imbalance  metric lies in its ability to accurately identify and distinguish between these two qualitatively different  ordered  states.

Furthermore, the nonmonotonic behavior of the imbalance metric in the WS small-world model reveals the complex intrinsic trade-off between  efficiency  and  fairness.  A local modification aimed at improving the efficiency of a highly ordered system may paradoxically lead to a severe disruption of its global functional fairness. Therefore, the comprehensive performance of a network is far from what a single structural or efficiency metric can summarize, and  Imbalance  provides an indispensable new perspective for examining this complex  structure-function  relationship.

\subsection{The Theoretical Positioning of  Imbalance  and the Significance of its Parameterized Design}
From a deeper perspective, the parameterized design of the  Imbalance  metric actually reveals that network  functional fairness  itself is not an absolute intrinsic property, but rather a relational property dependent on the observer (or specific functional requirements). A network is not  imbalanced  in itself; only under a specific  ruler  (the quality of service requirements defined by $h_0, a$) does its imbalanced state emerge and become quantifiable. Therefore,  Imbalance  is not just a measurement tool; it embodies a  context-dependent network evaluation paradigm . This paradigm acknowledges that, in dealing with complex and diverse real-world network problems, seeking an  adjustable lens  that can adapt to different scenarios and answer different questions is more realistic and scientifically valuable than pursuing a single universally  optimal  metric. For example, we can define $a$ and $h_0$ as follows:

Step 1: Define Service Level Agreement (SLA): Clearly define the SLA with which you are concerned. For example, for VoIP services, the SLA might be  99\% of path delays should be below $50ms$. 
Step 2: Map to Topological Parameters: Map the SLA to the parameters of the I metric. For example, if the average delay per hop in the network is $10ms$, then $h_0$ can be set justifiably to 50 ms / 10 ms = 5 hops. The parameter $a$ can reflect the tolerance for exceeding the SLA; the stricter the SLA, the larger the value of $a$.
Step 3: Perform Sensitivity Analysis: We emphasize that for a new network, the best practice is to plot the phase diagram of $I(h_0, a)$ , observe the sensitive regions of the I value to the parameters, and thus find the vulnerability points of the network in terms of functional fairness.

\subsection{Limitations and a Road map for Future Research}
As a brand-new theoretical framework, the exploration of the  Imbalance  metric is still in its nascent stage, and its applicability boundaries and future potential have yet to be discovered. The current work mainly has the following expandable directions:
\begin{itemize}
    \item \textbf{Theoretical Level}: The primary theoretical extension of the current work is to provide rigorous proofs for the core mathematical underpinnings of the Imbalance metric. This mainly includes two aspects: first, formalizing and proving the monotonic relationship between Imbalance and the variance of the all-pairs shortest-path distribution; second, establishing a global constraint relationship between Imbalance and the network s algebraic connectivity $\lambda_2$ (i.e., an upper bound determined by $\lambda_2$). These proofs will lay a more solid mathematical foundation for the theoretical edifice of Imbalance and will be the core of our next phase of work. Additionally, extending the current framework based on unweighted graph shortest hops to weighted, directed, time-varying, and multilayer networks is key to enhancing its universality.
    \item \textbf{Computational Level}: The  I  metric relies on APSP computation, and its $O(N^3)$ complexity limits its application on very large-scale networks. Developing efficient approximation algorithms or sampling-based estimation methods will be an important step towards its broader practical application.
    \item \textbf{Empirical Level}: Applying the I metric to various real-world networks (such as transportation networks, power grids, social networks, biological networks) is a promising research direction. A particularly valuable application would be in the analysis of financial networks, for instance, by applying it to inter-bank transaction data. The metric could help identify systemically important financial institutions and reveal critical pathways of financial contagion, thus providing a new quantitative tool for assessing financial stability. Exploring its practical application value in areas such as network resilience prediction and evolutionary mechanism analysis also remains an important avenue for future work.
\end{itemize}

\subsection{Conclusion}
In summary, the  imbalance   metric proposed in this paper provides a novel perspective to evaluate and understand complex networks, centered on functional fairness, which goes beyond traditional descriptions of efficiency and structure. It is not only an operational quantitative tool, but also a design philosophy that guides us in thinking about  what constitutes a better network.  We believe that this metric and its subsequent research will inject new theoretical vitality and practical value into the fields of network science, systems engineering, and the broader study of complex systems.

\begin{acknowledgments}

This work was supported by the National Key Research and Development Program of China under Grant No. 2024YFE0200300. The authors would also like to thank the anonymous reviewers for their insightful comments and valuable suggestions, which have greatly improved the quality of this paper.
\end{acknowledgments}

% --- BIBLIOGRAPHY ---
% The bibliography is generated from your Markdown file.
% I have created BibTeX keys like ref1, ref2, etc.
\bibliographystyle{apsrev4-2} % 使用标准的APS样式，通常是apsrev4-2.bst

\bibliography{reference}

\end{document}